\documentclass[twocolumn]{aastex701}
\usepackage{newtxtext,newtxmath}
\usepackage{graphicx}	% Including figure files
\usepackage{amsmath}	% Advanced maths commands
\usepackage{bm}		% Bold maths symbols, including upright Greek
\usepackage{color}
\usepackage{cancel}
\usepackage{pifont}

\newcommand{\bma}{\left(\begin{matrix}}
\newcommand{\ema}{\end{matrix}\right)}
\newcommand{\ba}{\begin{aligned}}
\newcommand{\ea}{\end{aligned}}

\newcommand{\D}{\text{d}}
\newcommand{\C}{\mathbb{C}}
\newcommand{\p}{\partial}
\newcommand{\fe}{\text{fe}}
\newcommand{\pe}{\text{pe}}

\newcommand{\bgeq}{\begin{equation}}
\newcommand{\edeq}{\end{equation}}
\newcommand{\braket}[1]{\left\langle #1 \right\rangle}
\newcommand{\dd}[2]{\frac{\mathrm{d}#1}{\mathrm{d}#2}}
\newcommand{\DD}[2]{\frac{\mathrm{D}#1}{\mathrm{D}#2}}
\newcommand{\pp}[2]{\frac{\partial #1}{\partial #2}}

\newcommand{\xkh}[1]{\left( #1 \right)}
\newcommand{\zkh}[1]{\left[ #1 \right]}
\newcommand{\dkh}[1]{\left\{ #1 \right\}}
\newcommand{\ini}{\text{ini}}
\newcommand{\fin}{\text{fin}}
\newcommand{\midd}{\text{mid}}
\newcommand{\MHD}{\text{MHD}}
\newcommand{\gPIC}{\text{gPIC}}
\newcommand{\sfP}{{\sf P}}
\newcommand{\sfT}{{\sf T}}

\begin{document}

\title{Magnetohydrodynamic-guiding-center-particle-in-cell Method for Multiscale Plasma Kinetic Simulations}%\footnote{Released on March, 1st, 2021}}

\author[0000-0001-5185-6114]{Zitao Hu}
\affiliation{Department of Astronomy, Tsinghua University,
Beijing 100084, China}
\email[show]{huzt@mails.tsinghua.edu.cn}  

\author[0000-0001-6906-9549]{Xue-Ning Bai}
\affiliation{Department of Astronomy, Tsinghua University,
Beijing 100084, China}
\affiliation{Institute for Advanced Study, Tsinghua University, 
Beijing 100084, China}
\email[show]{xbai@tsinghua.edu.cn}  

\author[0009-0009-7676-6188]{Xiaochen Sun}
\affiliation{Department of Astrophysical Sciences, 
Princeton University, Princeton, New Jersey 08544, USA}
\email{}

\begin{abstract}
We present the formulation, algorithm and numerical tests of the magnetohydrodynamic-particle-in-cell (MHD-PIC) method with particles treated under the guiding center approximation, which we term the MHD-gPIC method, and it is implemented in the \verb|Athena++| MHD code. The new MHD-gPIC model consists of thermal (cold) fluid and high-energy particles whose dynamics are integrated through guiding center equations including drift motion, with carefully evaluated source terms as particle backreaction. The code is validated with a series of tests, and it is expected to be primarily applicable to study particle acceleration and transport in systems where gyro-resonance is considered insignificant. We also present preliminary studies of particle acceleration during non-relativistic magnetic reconnection. 
\end{abstract}


\section{Introduction}
The universe is filled with high-energy particles that exhibit an exponential power-law energy spectrum, yet their origins and acceleration mechanisms remain a subject of intense research. 
Among the potential mechanisms, magnetic reconnection \citep{Guo2023} stands out as a likely candidate for efficient particle acceleration, supported by a wealth of observational evidence such as solar flares and other astrophysical phenomena.
The physics of particle acceleration necessitates a deep understanding of the microscopic processes that occur during reconnection. 
However, the initial scales at which particles are accelerated reside in the kinetic scales of the background plasma, which is vastly different from the macroscopic scales of the corresponding astrophysical system. {More specifically}, plasmoid instability {in current sheets has been well recognized as} a key mechanism that can lead to fast reconnection {\citep{Loureiro2007, Uzdensky2010}}. 
%It is a dynamical instability that occurs in current sheets, 
It leads to the formation of small-scale plasmoids, {followed by a hierarchical merging processes towards} macroscopic scales, which play
%These plasmoids can then merge, leading to a significant increase in the reconnection rate. 
%This process has been shown to be a possible pathway for fast reconnection, and it plays 
a crucial role in the acceleration of particles. 
%This scale separation presents a significant challenge for first-principles numerical simulations, which are crucial for elucidating the intricate dynamics of particle acceleration during reconnection.
%The plasmoid instability operates across multiple spatial scales, leading to a hierarchy of plasmoids that span from the kinetic to the fluid scales.
{In the context of solar flares, the typical background plasma scale (electron skin depth) is on the order of $\sim$cm, as opposed to system scale of $\sim10^{10}$cm \citep{Chen2020}.}
This huge scale separation poses significant challenges for numerical simulations, as it requires resolving the kinetic physics of particles while also capturing the fluid-scale dynamics.

Traditional magnetohydrodynamic (MHD) simulations are capable of capturing the large-scale dynamics of astrophysical phenomena but do not account for the detailed kinetic effects of non-thermal particles. 
On the other hand, particle-in-cell (PIC) simulations, which treat particles as individual entities, can resolve the microscopic scales but come with the significant computational cost associated with resolving the kinetic physics of particles (e.g., \citealt{Birdsall1991, Fedeli2022}).
A hybrid-PIC approach (e.g., \citealt{Lipatov2002, Kunz2014}), attempts to bridge this gap by treating electrons as a fluid while allowing for kinetic ions evolution. 
This method is computationally less expensive than full PIC but still requires a significant amount of computation to resolve the ion scale.
The MHD-PIC method is designed to overcome the limitations of these traditional approaches \citep{bai2015, Sun2023}. 
By treating the thermal plasma as a fluid described by MHD and only the non-thermal particles as kinetic particles, the MHD-PIC method allows for the resolution of large-scale dynamics while capturing the full kinetic nature of the non-thermal particles. 
This method only requires the resolution of particle gyration, and
%and back-reaction, which introduces computational complexity but also enables the study of detailed particle dynamics and feedback.
%The MHD-PIC method, with its ability to resolve both the macroscopic MHD dynamics and the microscopic particle physics, 
is particularly well-suited for simulating cosmic ray transport.
However, in the context of magnetic reconnection, {the gyro-radius of even the energetic particles are in general significantly smaller than the scales over which the magnetic field varies, and it has been found that particle acceleration is primarily through drift acceleration mechanisms, instead of resonant interactions with waves at gyro-scale \citep{Drake2006, Drake2013, Li2019}}.
%where a significant number of particles are accelerated within the reconnection region, there is a desire to further refine the model to capture the intricate physics of this process.

One potential approach to enhancing the MHD-PIC model for magnetic reconnection is to employ the guiding center approximation (GCA). 
%This approximation is particularly useful because in the reconnection process, the magnetic energy dominates, leading to particle gyration scales that are significantly smaller than the scales over which the magnetic field evolves. 
By treating particles as guiding centers (GCs), we can
{bypass the gyro-scale, thus further alleviating the issue of scale separation}
%reduce the computational complexity 
while still capturing the essential physics of particle acceleration. 
We note that similar ideas have been explored in other works.
In \citet{Drake2019}, \citet{Arnold2019}, and \citet{Yin2024}, the authors developed their \textit{kglobal} model and applied it to magnetic reconnection \citep{arnold2021}. 
Nevertheless, their {base MHD} code 
%requires a high-order resistivity, and the particles' share is fixed in their model. 
{relies on artificial diffusion for stability, and uses different equation of states for background (thermal) electrons and ions.}
In contrast, \citet{Mignone2023} added a guiding center approximation for relativistic particles to the \verb|PLUTO| code while particle feedback was not considered in their model.

In this paper, we introduce the MHD-gPIC model, which extends the MHD-PIC framework \citep{bai2015} by incorporating the guiding center approximation.
We re-derived the guiding center equations and carefully treated the feedback of particles on the plasma rigorously. 
%Also, particle injection is permitted and self-consistant in the MHD-gPIC model.
We have implemented the new MHD-gPIC framework based on the MHD-PIC module of \citep{Sun2023} built upon the grid-based Godunov MHD code \verb|Athena++| \citep{Stone2020}. {We have made substantial effort in mitigating numerical issues especially particle noise reduction, allowing smooth user-defined particle injection, and have} validated our implementation through a series of benchmark tests. 

This paper is organized as follows. 
We derive the formulation of the MHD-gPIC model in Section \ref{sec:method}. 
Implementation of this formulation to the \verb|Athena++| MHD code is described in Section \ref{sec:implementation}.
Benchmark tests are provided in Section \ref{sec:tests}. 
We summarize and conclude in Section \ref{sec:summary}.

\begin{table*}\label{tab:symbols}
\begin{center}
\begin{tabular}{lcc}
  \hline
   & variable  & symbol  \\
  \hline
  particle $k$ & charge, mass   & $q_k,\ m_k$\\
   & real/GC position     & $\bm x_k,\ \bm X_k$\\
   & real/GC velocity     & $\bm v_k,\ \bm V_k$\\
   & real/GC momentum     & $\bm p_k,\ \bm P_k$\\
   & Lorentz factor   & $\gamma_k$\\
   & gyro-frequency   &  $\omega_k$ \\
   & magnetic moment   & $\mu_k$ \\
  \hline
  MHD & magnetic field/direction & $\bm B,\ \bm b$\\
  & magnetic curvature & $\bm \kappa$\\
  & (ideal-MHD) electric field & $\bm E$, ($\bm E_0$)\\
  & fluid velocity & $\bm u$\\
  & fluid/MHD gas pressure & ${\sf P}_f,$ ${\sf P}$\\
  & fluid energy density & $\mathcal{E}_f$\\
  & MHD pressure/energy density & ${\sf P}^*,\ \mathcal{E}$\\
  & ratio of specific heats & $\gamma$\\
  \hline
  gPIC & momentum/energy feedback & $\bm F_\gPIC,\ W_\gPIC$ \\
  & mass ratio & $R$\\
  \hline
  general & number/mass density & $n,\ \rho$\\
  &bulk velocity (individual components) & $\bm u_{\rm p,fe,fi}$\\
  & charge/current density & $q,\ \bm J$\\
  & speed of light & $\C$\\
  & pressure/stress tensor (scalar) & $\mathbf P$ ($\sfP$), $\mathbf T$ ($\sfT$)\\
  \hline
  time derivative & in observer's frame & $\pp{} t$\\
  & in MHD gas co-moving frame & $\DD{}t\equiv\pp{}t+\bm u\cdot\nabla $\\ 
  & in particle k's co-moving frame & $\dd{}t\equiv \pp{}t+\bm V_k\cdot\nabla$\\ 
  \hline
  subscripts & parallel/perpendicular-$\bm b$  &  $_\|$, $_\perp$\\
  & fluid total/ion/electron & $_f$, $_\text{fi}$, $_\fe$ \\
  & gPIC total/ion/electron & $_p$, $_\text{pi}$, $_\pe$\\
  & total ion/electron & $_i$, $_e$\\
%   & virtual electron & $\tilde{\ \ }_{\fe}$\\
  \hline
\end{tabular}
\end{center}
\caption{A list of symbols used in this paper.
%{[You use $u$ for both bulk velocity and fluid velocity. Though I'm not sure what's a better way to avoid overlap.]}}{[yes, $\sfP$ and $\bm u$ are used for both gPIC and MHD part, the difference is there will be a subscript when used for gPIC while no subscript for MHD.]{[How about this? (I'm experimenting.)]}
}
\end{table*}

\section{Methods}\label{sec:method}
In our MHD-gPIC model, we simulate the motion of charged particles (both ions and electrons) %, specifically electrons,
in magnetic fields and their dynamic interaction with the surrounding thermal plasma at a macroscopic scale that greatly exceeds the scale of their gyro-radii.
The model is structured into two primary components: the equations of motion for the non-thermal components, %electrons 
which are treated as discrete particles, and the MHD equations that govern the behavior of the ambient thermal gas.
The gas is considered to be non-relativistic. Similar to the MHD-PIC model \citep{bai2015, Sun2023}, the particles can be non-relativistic or relativistic, and in the latter case, particle velocity is limited by an artificial speed of light $\mathbb{C}$, and the formalism is valid as long as $\mathbb{C}$ well exceeds any MHD velocities.
%To represent the non-thermal particles, %electrons, 
%we adopt the methodology used in 
Similar to the standard PIC and MHD-PIC
models, non-thermal particles are considered as super particles that effectively stand for a large number of physical particles, %electrons. 
carrying certain effective shape matching the simulation grids to facilitate interpolations and deposits.
%To ensure overall charge neutrality within the MHD part of the model, 

Our formalism and its derivation inevitably involve a large set of symbols, which we summarize in Table \ref{tab:symbols}. 
We will introduce the relevant notations as we delve into the derivation. In brief, for particles, we distinguish real and guiding-center averaged quantities in lower-case and upper-case letters, respectively. 
In the fluid equations including the fluid-particle interface, we use subscript `$_f$' and `$_p$' to denote fluid and particle quantities (where fluid quantities are by definition assumed to be thermal),  
within which we further distinguish between contributions from ions and electrons by subscripts `$_i$' and `$_e$'. 
% In addition, contributions from virtual particles {(to be introduced next)} are denoted with a `$\tilde{\ \ }$'. 
We use $\perp$ and $\|$ to denote vector components perpendicular and parallel to magnetic field, respectively. We also note that we use $\bm u$ to denote fluid velocities, while use $\bm v$ to denote particle velocities.

We first introduce the main framework and formulation, before delving in to a more detailed derivation for particle integration and feedback through Sections \ref{sec:particle} to \ref{sec:gas_dynamics}.
For a particle labeled $k$ with charge $q_k$, mass $m_k$ and momentum $p_k$, its magnetic moment
%we postulate that its magnetic momentum, 
$\mu_k=p_{k\perp}^2/2m_kB$ remains constant throughout its trajectory under the guiding-center approximation.
%where we use $\perp$ and $\|$ to define components perpendicular and parallel to magnetic field, respectively.
The parallel motion of this particle is governed by {equation (\ref{eq:v_gk}) and}
\begin{equation}
    \dd{P_{k\|}}t=\Delta v_{k\|}\bm P_{k}\cdot\bm \kappa+\gamma_k\bm u_{\perp}\cdot\nabla_\|\bm u-\frac{\mu_k}{\gamma_k}\nabla_\|B+q_kE_\|,\label{eq:particle_momentum}
\end{equation}
where the $\bm u$ denotes the MHD gas velocity. 
The term $\Delta v_{k\|}\equiv v_{k\|}-u_{\|}$ represents the particle’s parallel velocity in the gas co-moving frame.
The vector $\bm P_{k}$ is the gyro-averaged momentum of the particle, while $\bm\kappa=\bm b\cdot\nabla\bm b$ represents the magnetic curvature.
Here, $\bm b$ denotes the direction of the magnetic field, $\gamma_k$ is the particle Lorentz factor, $\nabla_\|\equiv \bm b\cdot\nabla$ is the gradient operator parallel to $\bm b$ and $E_\|$ denotes the parallel electric field, to be given {by equation (\ref{eq:epara1})} in Section \ref{ssec:Epara}.

The dynamical equations governing the MHD component, written in conservative form with source terms from the gPIC part, read
\begin{align}
    &\pp{\rho_f}t+\nabla\cdot(\rho_f\bm u)=0,\\
    &\pp{\rho_f\bm u}t+\nabla\cdot\xkh{\rho_f\bm{uu}-\frac{\bm {BB}}{4\pi}+\sf{P}^*}=-\bm F_\gPIC,\label{eq:momentum_feedback_predict}\\
    &\pp{\mathcal{E}}t+\nabla\cdot\zkh{({\sf P}^*+\mathcal{E})\bm u-\frac{\bm {BB}\cdot\bm u}{4\pi}+\frac c{4\pi}(\bm{E}-\bm E_0)\times\bm B}=-W_\gPIC,\label{eq:energy_feedback_predict}
\end{align}
where the symbols have their usual meanings (summarized in Table \ref{tab:symbols}), {$\bm F_\gPIC$ and $W_\gPIC$ are given by equation (\ref{eq:F_gPIC}) and equation (\ref{eq:W_gPIC}), respectively}. In particular, $E_0=\bm B\times\bm u/\C$ is the ideal MHD electric field, ${\sf P}^*\equiv {\sf P}+B^2/8\pi$ is the MHD pressure with ${\sf P}={\sf P}_f+{\sf P}_{p\perp}$ being the {MHD} fluid pressure, 
% {[This may not be accurate when you also have ion particles. Please correct/clarify]}
the total energy density
\begin{equation}
    \mathcal E\equiv\mathcal E_f+\frac{B^2}{8\pi}=\frac{{\sf P}_f}{\gamma-1}+\frac12\rho u^2+\frac{B^2}{8\pi}.\label{eq:total_E}
\end{equation}
{where we also define the fluid energy density $\mathcal E_f$ which does not contain contribution from magnetic field.}

%{In order to mitigate particle Poisson noise, the MHD pressure is formulated by incorporating the perpendicular pressure $\sfP_{p\perp}$ from the gPIC part, while the MHD energy remains unchanged.}
Our definition of MHD pressure differs from the standard definition by incorporating the perpendicular pressure $\sfP_{p\perp}$ from the gPIC part. The reason behind this definition is as follows.
In the presence of non-thermal particles, charge neutrality of the composite system demands $n_{\rm fe}+n_{\rm pe}=n_{\rm fi}+n_{\rm pi}$. The number densities of electrons and ions in the background thermal fluid are generally non-equal ($n_{\rm fe}\neq n_{\rm fi}$) , while they are considered to share the same temperature $T$. With electrons being essentially massless, they only contribute to pressure but not mass. As a result, the total fluid pressure is
\begin{equation}
{\sf P}_{f}={\sf P}_{\rm fe}+{\sf P}_{\rm fi}\propto n_{\rm f}T,
\end{equation}
where $n_{\rm f}=n_{\rm fe}+n_{\rm fi}$ {is the total particle number density in the MHD part}. 
{Generally, gas temperature $T$ and total plasma particle number density $n_f+n_p$ should be smooth in the simulation.}
{However, particle distribution $n_p$ in the PIC part is generally not smooth due to bulk motion or local particle injection.}
%If PIC particle number is not high enough, fluctuation would be significant.
%Furthermore, initial distribution and injection method will also induce abnormal local distribution of PIC particles.
% However, the variation in $n_{\rm f}$ makes the fluid part incompatible with the standard ideal gas equation of state (${\sf P}_f\propto\rho_fT$) nearly ubiquitously adopted in astrophysical MHD, 
This makes it particularly inconvenient for computational purposes {since the related gradient computation in equation (\ref{eq:momentum_feedback_predict}) and (\ref{eq:energy_feedback_predict}) will amplify the noise at grid scale}.
{As a partial remedy}, we add the perpendicular-$\bm B$ pressure of the gPIC part ${\sf P}_{p\perp}$ to the gas thermal pressure.
In this way, we have ${\sf P}\equiv{\sf P}_f+{\sf P}_{p\perp}$ for total fluid pressure, {i.e., the total pressure of the plasma by assuming parallel-$\bm B$ pressure of the PIC part is the same as its perpendicular pressure}. 
The advantage of this trick is that it can effectively mitigate numerical noise generated by particle feedback ({as we shall see} in section \ref{sec:virtual_electron}), thereby enhancing the stability of our numerical scheme.

%The physical significance of these equations, along with 
The terms on the right hand side of Equations (\ref{eq:momentum_feedback_predict}) and (\ref{eq:energy_feedback_predict}), $\bm F_\gPIC$ and $\ W_\gPIC$, represent the backreaction from the non-thermal particles (gPIC part) to the thermal fluid. These expressions and derivations will be elaborated in section \ref{sec:gas_dynamics}. We note that in the standard MHD-PIC formalism, the backreaction is simply treated by ensuring momentum and energy conservation following detailed derivations \citep{bai2015}. The derivation there assumed the number density of non-thermal particles is much smaller than that of the background gas so that {$n_{p}\ll n_i$}. 
Our formalism for the backreaction is compatible with that in standard MHD-PIC, but further relaxes the assumption of small particle number density. This leads to a revision to the gas pressure and hence the entire derivation, and we will delve into the details in subsequent subsections. 
% We further demonstrate the robustness and performance of the virtual particle approach through a comparative benchmark test in Appendix \ref{appendix:virtual_electron}.
Furthermore, the derivation in this section can be generalized to fully general-relativistic MHD in curved spacetimes, and we leave this part in Appendix \ref{appendix:GRMHD-gPIC}.
Additionally, we have included some intermediate steps of the derivations in Appendix \ref{appendix:deduction}.

\subsection{Particle Equarion of Motion}\label{sec:particle}
For a single particle labeled $k$, either an electron or an ion, we adopt the guiding center approximation (GCA), which replaces the particle’s actual trajectory with its center of motion. 
This approach is justified by the fact that the particle’s gyroradius and gyro-timescale is significantly smaller than the corresponding characteristic scales of the MHD system. 
The guiding center approximation of particle motion has been widely studied since \citet{northrop1963}, with relativistic representations subsequently derived \citep{Mignone2023, Trent2024}.
We list the main derivation in appendix \ref{appendix:GC_equations}, where we provide the most general derivation in covariant form. In the main text below, we rewrite the results in the more familiar $3+1$ form.
%and use the results directly in this section. 
%{[Should comment on the difference from the more recent literature, e.g., Mignone+23.]}

We use two sets of variables to distinguish primitive and gyro-averaged quantities for particles, where the former is denoted by lower-case letters, while the latter is represented by the corresponding upper-case letters.
%To represent gyro-averaged quantities associated with the guiding center, we employ the corresponding capital letters. %subscription “$_g$”. 
For instance, the location of particle $k$'s guiding center is
\begin{equation}
    \bm X_{k}\equiv\bm x_k+\frac{1}{\omega_kB}(\bm v_k\times\bm B+\bm E), \label{eq:x_gk}
\end{equation}
%$\bm b$ is the unit vector aligned with the magnetic field direction 
%and 
where $\omega_k=q_kB/\gamma_km_k$ is the gyro-frequency of the particle.
The guiding center velocity for the particle electron, denoted as  $\bm V_{k}$, is given by
\begin{equation}
    \bm V_{k}\equiv\dd{\bm X_{k}}t=v_{k\|}\bm b+\bm u_{\perp}+\bm v_\text{drift},\label{eq:v_gk}
    % \bm v_{gk}\equiv\dd{\bm x_{gk}}t=v_{k\|}\bm b+\bm v_{\bm E,k}+\underbrace{\frac{\mu_k J_\|}{\omega_kB}\bm b}_{\bm v_{s,k}}+\underbrace{\frac{\mu_k}{\omega_kB}\bm b\times\nabla_\|B}_{\bm v_{\nabla B,k}}+\underbrace{\frac1{}\bm b\times\bm\kappa}_{\bm v_{\bm\kappa,k}},\label{eq:v_gk}
\end{equation}
where $\bm u_{\perp}\equiv\bm E\times\bm Bc/B^2$ is the perpendicular MHD gas velocity (a.k.a. $\bm E\times\bm B$ drift velocity),
% where we omit the high frequency terms.
%On the right hand side of Equation (\ref{eq:v_gk}), 
$v_{\|k}$ is the parallel the particle velocity, $\bm v_\text{drift}$ is the
%{$\mathscr{O}(\omega_k^{-1})$}
% first order drift 
perpendicular velocity of the guiding center relative to the fluid (see Equation (\ref{eq:GC_V1})).
% {[Please give the full expression.]}
% {Among these terms, $\bm v_\text{drift}$ of the order $\mathscr{O}(\omega_k^{-1})$ relative to [??? Please clarify.]}
We keep the leading order $\mathscr{O}(\omega_k^{-1})$ terms which corresponds to Speiser motion, curvature drift, {inertial drift} and gradient-$B$ drift
\begin{equation}
    \bm v_\text{drfit} = \frac1{\gamma_k\omega_k}\zkh{\underbrace{\frac{\mu_kJ_\|}{m}\bm b}_{\rm Speiser}-\xkh{\underbrace{v_{k\|}^2\bm \kappa}_{\rm curvature}+\underbrace{\gamma_kv_{k\|}\DD{\bm b}t}_{\rm inertial}+\underbrace{\frac{\mu_k}{m}\nabla B}_{\text{grad-}B}}\times\bm b},\label{eq:v_drift}
\end{equation}
%In GCA, which is primarily applicable 
where $J_\|=(\nabla\times{\bm B})_\|$ is the parallel current density of plasma.
%{[So this is $(\nabla\times{\bm B})_\|$?]}{[Yes]}
As the GCA applies when the gyro-timescale $\omega_k^{-1}$ is significantly smaller than the time scale corresponding to the variations in the magnetic field experienced by the particle, we only use the other two $\mathscr{O}(1)$ terms in most cases, while provide the option to retain the above $\bm v_\text{drift}$ terms in the code.

The momentum and energy equations (i.e., equation (\ref{eq:GC_ME}) and (\ref{eq:GC_E}) in observer's coordinate with $\D t=\gamma_k\D\tau$) in GCA {are}
\begin{align}
    &\dd{\bm P_{k}}t=q_k\xkh{\bm E+\bm V_{k}\times\bm B}-\frac{\mu_k}{\gamma_k}\nabla B,\label{eq:electron_momentum_1}\\
    &\dd{\varepsilon_{k}}t=q_kE_\|V_{k\|}+\frac{\mu_k}{\gamma_k}\xkh{\DD{B}t-u_\|\nabla_\|B}\nonumber\\
    &\quad+\gamma_km_k\xkh{V_{k\|}\Delta V_{k\|}\bm u\cdot\bm \kappa+V_{k\|}\bm u\cdot\DD{\bm b}t+ u_\perp\dd{u_\perp}t},\label{eq:electron_energy}
\end{align}
where $\bm P_{k}\equiv\gamma_km_k\bm V_{k}$ denotes the gyro-averaged momentum, $\gamma_k=\sqrt{1+P_k^2/m_k^2\C^2+2\mu_kB/m_k\C^2}$ is the Lorentz factor and $\mu_k\equiv{p_{k,\perp}^2}/{2m_kB}$ is the magnetic moment which is an adiabatic invariant {and $\Delta \bm V_k\equiv\bm V_k-\bm u$ is the particle velocity in gas co-moving frame}.
We also defined $\D/\D t\equiv \p/\p t+\bm V_k\cdot\nabla$ and $\mathrm{D}/\mathrm{D}t\equiv\p/\p t+\bm u\cdot\nabla$ when they act on MHD variables.
%Since $v_\text{drift}\ll v_{E}$ in guiding center approximation, only the parallel momentum matters for each particle.
{The energy equation will be used to derive the energy feedback of gPIC part in section \ref{ssec:particle_feedback}.}

We can decompose {the momentum equation} into two distinct components: one aligned with the magnetic field and the other orthogonal to it.
These components are expressed as
\begin{align}
    \dd{P_{k\|}}t=&\dd{P_k}t\cdot\bm b+\bm P_{k}\cdot\dd{\bm b}t=\bm P_{k}\cdot\dd{\bm b}t+qE_\|-\frac{\mu_k}{\gamma_k}\nabla_\|B,\label{eq:electron_momentum_0}\\ 
    \dd{\mu_k}t=&\ 0.
\end{align}
In the parallel momentum equation (\ref{eq:electron_momentum_0}), the first term on the right-hand side originates from the variation of the magnetic field direction over time. %{[What do you mean?]}
%represents first-order Fermi acceleration 
%{(please explain further. Main contribution from vE and vdrift?)}.
This term is typically treated with the approximation $\D\bm b/\D t\approx \bm V_k\cdot\nabla$
%{I think no, in the past they use $v_K\cdot\nabla$.} 
in previous studies (e.g. \citealt{northrop1963}, \citealt{Drake2019}), which is valid when the particle speed $V_k$ is much greater than the MHD speed $u$. However, we stress that even this condition is satisfied, the approach violates Galilean invariance, leading to different guiding-center orbits in different reference frames.
%{(explain what $d/dt$ and $\p/\p t$ mean.)}

Instead, we transform the rate of change of the magnetic field direction from the particle’s frame to the MHD frame using:
\begin{equation}
    \dd{\bm b}t=\Delta V_{k\|}\bm \kappa+\DD{\bm b}t,\label{eq:dbdt}
\end{equation}
where we have only kept the zeroth order velocities.
%{[I get confused again. What is the difference between $\bm v_{\rm MHD}$ and $\bm u$?]}{[sorry, I missed them when I was changing notations earlier.]}
By combining (\ref{eq:electron_momentum_0}) with (\ref{eq:dbdt}), we obtain the parallel momentum equation:
\begin{equation}
    \dd{P_{k\|}}t=\Delta V_{k\|}\bm P_{k}\cdot\bm \kappa+\bm P_{k}\cdot\DD{\bm b}t-\frac{\mu_k}{\gamma_k}\nabla_\|B+qE_\|.\label{eq:electron_momentum}
\end{equation}
In ideal MHD, the second term on the right-hand side of (\ref{eq:electron_momentum}) is:
\begin{equation}
    \DD{\bm b}t=(1-\bm{bb})\cdot\nabla_\|\bm u\ .\label{eq:dbdt_MHD}
\end{equation}
%which is utilized in the particle’s momentum equation when electric field from other source terms terms, such as Ohmic resistivity, is unimportant. 
%{We keep this term in our equations.}
% {[It is a little confusing: do you keep this term or not?]}
%{(I don't understand how this is derived. Also, this is a pure MHD thing, why do you mention particles in the explanation?)}
By substituting (\ref{eq:v_gk}) and (\ref{eq:dbdt_MHD}) in (\ref{eq:electron_momentum}),
%into the appropriate places, 
we derive the full particle momentum equation (\ref{eq:particle_momentum}).
%{(I again don't understand how this is done.}
Our treatment by the usage of (\ref{eq:dbdt}) preserves Galilean invariance, as we demonstrate in a test problem in Section \ref{sec:loop}.

For the remaining two terms in equation (\ref{eq:electron_momentum}), the third term corresponds to betatron acceleration, as it arises from the conservation of magnetic moment in the perpendicular equation. 
The final term is attributed to the non-ideal parallel electric field, to be specified in Section \ref{ssec:Epara}.

\subsection{Particle-Fluid Interface}\label{sec:fluid_closure}
%To establish a link between the evolution of particle components and the MHD framework, we must account for the fluid closure equations that encapsulate the comprehensive dynamics of the super particles. 
%This integration is achieved by adhering to the principle of kinetic energy conservation, thereby integrating the feedback from the particle component into the MHD evolution equation.
In this subsection, we lay out the framework for the particle-fluid interface, so that we can estimate the momentum and energy exchange between the particle components with the thermal gas.
Inherited from the PIC approach, each particle carries an effective shape designed to facilitate interpolation from the grid, i.e., the macroscopic variable $G$ at location $\bm x$ is linked with microscopic variable $g_k$ of particles by
\begin{equation}
    G(\bm x)\equiv\sum_k S(\bm x,\bm x_k)g_k,\label{eq:interface}
\end{equation}
where $S(\bm x,\bm x_k)$ is the shape function which follows the standard triangular-shaped cloud (TSC) scheme in our code.
For convenience, we use the Dirac-$\delta$ function $\delta(\bm x-\bm x_k)$ as the shape function which is intuitive in the following derivations.

With equation (\ref{eq:interface}), we can get the macroscopic parameters of the gPIC part.
The number density $n_p$, mass density $\rho_p$ and charge density $q_p$ of particles are
\begin{align}
    n_p(\bm x)&\equiv\sum_k\delta(\bm x-\bm x_k),\\
    \rho_p(\bm x)&\equiv\sum_k\gamma_km_k\delta(\bm x-\bm x_k),\\
    q_p(\bm x)&\equiv\sum_kq_k\delta(\bm x-\bm x_k).
\end{align}
%{[Why do you have $\gamma$ factors here?]}{[it should be $\sum_k\gamma_k\delta(\gamma_k(\bm x-\bm x_k))$, but in code, the effective shape for all particles is same...]}
Similarly, the bulk velocity and current density are
\begin{align}
    \bm u_p(\bm x) &\equiv \sum_k\gamma_km_k\bm v_k\delta(\bm x-\bm x_k)/\rho_p,\label{eq:macro_u}\\
    \bm J_p(\bm x) &\equiv \sum_k q_k\bm v_k\delta(\bm x-\bm x_k).\label{eq:macro_j}
\end{align}
We also use the particle stress tensor $\mathbf{T}_p$ in the following sections, which is defined by
\begin{equation}
    \mathbf{T}_p(\bm x)\equiv\sum_k\gamma_km_k\bm v_k\bm v_k\delta(\bm x-\bm x_k),
\end{equation}
and the particle pressure tensor $\mathbf{P}_p$, which is the stress tensor in particles' co-moving frame,
\begin{equation}
    \mathbf{P}_p\equiv \sum_k\gamma_km_k(\bm v_k-\overline{\bm u}_p)(\bm v_k-\overline{\bm u}_p)\delta(\bm x-\bm x_k)=\mathbf{T}_p-\rho_p\bm u_p\bm u_p.
\end{equation}
%{[No, it's in particle bulk motion frame, $u_p$ is defined in (22)]}
The above definitions apply to all particles (without additional subscripts) or to each type of particle (with additional subscript `$_e$' or `$_i$').

\subsection{Non-ideal Electric Field}\label{ssec:Epara}
%To complete the particle equation of motion, we follow the approach of \citet{Arnold2019} to derive the non-ideal parallel electric field.
%This electric field is employed to capture the acceleration of non-thermal particles, while the MHD component possesses its own non-ideal electric field, based on different physical assumptions and purposes.
%Therefore, the electric field introduced in this section is excluded from the induction equation, thereby avoiding the introduction of additional noise.
In this subsection, we complete the derivation of parallel electric field needed for particle motion (\ref{eq:particle_momentum}). Follow \citet{Arnold2019}, 
%For the non-ideal electric field, 
this can be determined using the generalized Ohm’s law:
\begin{equation}
    \bm E=-\frac{\bm u_i}c\times\bm B+\frac{\bm R_e}{en_e}+\frac1{en_e}\frac{\bm J\times\bm B}c-\frac1{en_e}\nabla\cdot\mathbf P_e-\frac{m_e}e\dd{\bm u_e}t,
\end{equation}
where $\bm u_i$ is the ion velocity, %equivalent to the MHD gas velocity $\bm u$, 
$\bm R_e$ represents the
%collision term 
frictional force between ions and electrons in the MHD gas, $n_e$ is the total electron number density, $\mathbf P_e$ is the total electron pressure tensor, and $\bm u_e$ is the averaged electron velocity.
%{(Be specific on whether non-thermal particles contribute to $n_e$ and $P_e$.)}
The last term in the equation can be disregarded 
%as $m_e/e\ll 1$.
in MHD scales given electron's small inertia.
For the parallel component $E_\|$, the second term is usually significantly smaller than the fourth (pressure gradient) term under typical collisionless conditions,
%in regard of the low plasma density in astronomy, 
leading to
%{[Haven't changed the symbols to $\sf P$ in this subsection?]}{[I use $\mathbf P$ for tensor and $\sfP$ for scalar, should I use $\sfP$ for both?]{[I see. Good with me.]}}
\begin{equation}
    E_\|\approx-\frac1{en_e}\bm b\cdot\nabla\cdot \mathbf P_e .\label{eq:epara}
\end{equation}
%this result aligns with \cite{Arnold2019}.
We see that $E_{\parallel}$ is primarily governed by the electron pressure gradient, and involves both fluid (thermal) and particle (non-thermal) electrons.

To compute $\mathbf{P}_{\text{e}}$, we write
\begin{equation}
    \mathbf P_e=\mathbf T_{\fe} +\mathbf T_{\pe}-\rho_e\bm u_e\bm u_e=\mathbf P_{\fe} +\mathbf T_{\pe}+\rho_{\fe}\bm u_\fe\bm u_\fe-\rho_e\bm u_e\bm u_e,
\end{equation}
where $\mathbf T_\fe$ and $\mathbf T_\pe$ are the stress tensors of gPIC electrons and MHD electrons respectively, $\rho_e$ is the mass density of all electrons.
% {
% To estimate the result of above equation, we consider two limits: $\rho_\fe\gg\rho_\pe$ and $\rho_\fe\ll\rho_\pe$.
% If $\rho_\fe\gg\rho_\pe$, the system is nearly MHD, which means $\rho_e\bm u_e\bm u_e\ll\mathbf P_e$.
% In this case, using $\rho_e\bm u_e=\rho_\fe\bm u_\fe+\rho_\pe\bm u_\pe$, we can find $\bm u_\fe\approx-\bm p_\pe/\rho_fe$ where $\bm p_\pe$ is particle electron momentum density.
% Therefore,
% \begin{equation}
%     \mathbf P_e\approx \mathbf P_\fe+\mathbf T_\fe+\frac{\bm p_\pe\bm p_\pe}{\rho_\fe}.
% \end{equation}
% In the other limit, $\rho_\fe\gg\rho_\pe$
% }
%Using equation (\ref{eq:macro_u}) for gPIC electrons and MHD electrons respectively
% $\mathbf T_\fe=P_\fe\mathbf I + \rho_\fe\bm u_\fe\bm u_\fe$ and $\mathbf T_\pe=\mathbf P_\pe + \rho_\pe\bm u_\pe\bm u_\pe$ where $\bm u_\fe$ and $\bm u_\pe$ are the bulk velocities of MHD electrons and gPIC electrons respectively, 
Using $\rho_e\bm u_e=\rho_\fe\bm u_\fe+\rho_\pe\bm u_\pe$ in above equation, with $\bm u_\pe$ from Equation (\ref{eq:macro_u}) for for gPIC electrons, we obtain
\begin{equation}
    \mathbf P_e=\mathbf P_\fe+\mathbf P_\pe+\rho_\fe\Delta\bm u_\fe\Delta \bm u_\fe+\rho_\pe\Delta\bm u_\pe\Delta \bm u_\pe,\label{eq:P_e}
\end{equation}
where $\Delta\bm u_\fe\equiv\bm u_\fe-\bm u_e$ and $\Delta\bm u_\pe\equiv\bm u_\pe-\bm u_e$.
Since particles have the same perpendicular velocity as the fluid part (ignoring the high-order drift velocities), we have $\Delta\bm u_\fe=-\rho_\pe\Delta\bm u_\pe/\rho_\text{fe}=\Delta u_{\fe,\|}\bm b$. 
% {[Because $u_e$ is a weighted average, this relation holds only when $\rho_{fe}=\rho_{pe}$.]}
%Equation (\ref{eq:P_e}), together with 
The pressure of fluid electrons ${\sf P}_{\fe}$ is given by %{(generalize this to be compatible with the inclusion of particle ions.)}
\begin{equation}
    {\sf P}_{\fe} =\frac{n_{\fe}}{n_\text{fi}+n_\fe}{\sf P}_f
    % =\frac{n_\text{i}-n_{\pe}}{2n_\text{i}}\frac{2n_\text{i}}{n_\text{fi}+n_\fe}{\sf P}_f=\frac{n_\text{i}-n_{\pe}}{2n_\text{i}}{\sf P}
    ,\label{eq:P_fe}
\end{equation}
where ${\sf P}_f$
%\equiv {2n_\text{i}}{\sf P}_f/\xkh{n_\text{fi}+n_\fe}$, i.e., 
is the gas pressure.
%{[Our symbol system has $P_f$, but what is $P_{\rm gas}$? Some additional equations on number conservations and definitions of various versions of P (especially related to how you use virtual particles) are needed to better understand Eq. 18.]}
%and we have assumed that fluid electrons share the same temperature as ions. 
Together with the pressure tensor of particle electrons $\mathbf P_{\pe}={\sf P}_{\pe,\|}\bm{bb}+{\sf P}_{\pe,\perp}(\mathbf I-\bm{bb})$, the final expression for the parallel electric field is given by
\begin{equation}
    E_\|=-\frac1{en_e}\zkh{\nabla_\|\xkh{{\sf P}_{\text{pe},\|}+{\sf P}_\text{fe}-\rho_e\Delta u_{\fe,\|}\Delta u_{\pe,\|}}-\Delta {\sf P}_{\text{e}}\nabla_\|\ln B},\label{eq:epara1}
\end{equation}
where $\Delta{\sf P}_{\text{e}}={\sf P}_{\text{pe},\|}-{\sf P}_{\text{pe},\perp}-\rho_e\Delta u_{\fe,\|}\Delta u_{\pe,\|}$ represents the electron pressure anisotropy. %, with $\Delta u_{\fe,\|}$ to be given in equation (\ref{eq:delta_u_fe}).

The remaining task is to calculate $\Delta u_\fe$. We start from total total current equation 
\begin{equation}
q_f\bm u-n_\fe e\Delta\bm u_\fe+\bm J_p= \frac{\C}{4\pi}\nabla\times\bm B, 
\end{equation}
%{where we assumed $\bm J_\fe\approx-n_\fe e \bm u_\fe$.}
%{[I thought this is obvious, but are you saying you have to make some approximation? Ignoring $\Delta u_{fi}$?]}
{where we have assumed that $\Delta\bm u_\fe$ obtained from momentum averaging also applies to velocity averaging (valid for non-relativistic fluids).}
{The} second term on the left-hand side signifies the relocation of fluid electrons to maintain charge neutrality. We can derive
\begin{equation}
    \Delta \bm u_\fe = \frac{n_\text{fi}-n_{\fe}}{n_\fe}u_\|\bm b+\frac{J_{p\|}\bm b}{n_\fe e}-\xcancel{\frac{\C\nabla\times\bm B}{4\pi n_\fe e}},\label{eq:delta_u_fe}
    % \Delta \bm u_\fe = \frac{\bm J_p-q_p\bm u-\nabla\times \bm B}{n_\fe e}\approx\frac{\bm J_p-q_p\bm u}{n_\fe e}\approx\frac{J_{p\|}-q_p u_\|}{n_\fe e}\bm b.
\end{equation}
where we have used $q_f=(n_\text{fi}-n_\fe)e$ and $\bm J_p\approx J_{p\|}\bm b-q_f\bm u_\perp$.
\footnote{Numerically, this result becomes unreliable when $|\Delta \bm u_{\fe}|\gg |\bm u_\fe|$.
This scenario can arise in regions where $n_\fe\ll n_\pe$. We will address how to handle this issue in the implementation section (section \ref{sec:implementation}).}
We omit the last (Hall) term in the above equation as it becomes important only on ion-inertial scales or below.
%for the same reason as in the deduction of the MHD equations. 
Note that if the particle electron pressure is isotropic, the second term in equation (\ref{eq:epara1}) equals zero, and the parallel electric force exactly balances the electron pressure gradient.

For our primary target application in magnetic reconnection, the non-ideal electric field may play a crucial role in particle injection and initial acceleration within the diffusion region near magnetic null points—such as X-points.
%in magnetic reconnection.
This effect has primarily been investigated in PIC simulations of relativistic reconnection \citep{French2023, Zhang2023}, where it has also been shown to enhance the reconnection rate \citep[e.g.,][]{Goodbred2022}.
However, the spatial scale at which this electric field is most prominent is typically on the order of the ion inertial length or smaller, whereas our typical applications target much larger (MHD) scales.
On these MHD scales, the non-ideal electric field—derived from the electron pressure gradient—primarily heats the thermal plasma rather than generating non-thermal particles \citep{arnold2021}.
Consequently, we include the option to disable this term. In particular, since its evaluation requires computing spatial gradients of particle quantities and their products, retaining it would otherwise introduce significant Poisson noise.
%although we treat it carefully in our code.
%{(Further comment on when this $E_\p$ is used and when you want to turn it off.)}
%{[Like the CR-induced Hall term in MHD-PIC, is there anything other than this $E_\parallel$ to modify the induction equation? Yes but very insignificant. Add a note.]}
We also note that this electric field is not applied to the MHD induction equation, where we apply standard resistivity to trigger reconnection following standard approaches.

\subsection{Particle Feedback}\label{ssec:particle_feedback}
{Once the particle equation of motion is established, the feedback from the gPIC component can be consistently derived.
A key distinction between the present model and the conventional MHD-PIC framework lies in the fact that the guiding center position does not coincide exactly with the particle’s physical position. 
This discrepancy necessitates careful treatment in the coupling between particle dynamics and the background gas.}

For each particle, base on equation (\ref{eq:x_gk}), the connection between its actual position $\bm x_k$ and its guiding center position $\bm X_{k}$ is encapsulated by the relationship
\begin{equation}
    \bm x_k = \bm X_{k} + \delta \bm x_k,\label{eq:particle_location}
\end{equation}
where the relative position $\delta \bm x_k$
%satisfies the gyro-phase averaging condition 
evolves as
$\D\delta\bm x_k/\D t=\omega_k\bm b\times\delta\bm x_k$ and
%the Larmor radius-related condition
its autocorrelation satisfies
$\braket{\delta \bm x_k\delta\bm x_k}=r_{k}^2(1-\bm{bb})/2$, where $r_{k}$ is the Larmor radius.
The force acting on the fluid
%closure of 
from the particles is
\begin{equation}
    \bm F(\bm x,t)\equiv\sum_k\dd{\bm p_{k}}t\delta(\bm x-\bm x_{k}),\label{eq:gPIC_momentum}
    % \rho_\gPIC\dd{\bm v_\text{gPIC}}t(\bm x,t)\equiv\sum_k\dd{\bm p_{k}}t\delta(\bm x-\bm x_{k}),\label{eq:gPIC_momentum}
\end{equation}
where $\text{d}\bm p_k/\text{d}t$ corresponds to the Lorentz force experienced by individual particles.
Similarly, the work done on the fluid %closure
from the particles is
\begin{equation}
    W(\bm x,t)\equiv\sum_k{\dd{\varepsilon_{k}}t}\delta(\bm x-\bm x_{k}),
\end{equation}
where $\varepsilon_k\equiv\gamma_km_k\C^2$ represents the energy of particle $k$. %{(Let's distinguish $c$ and $\mathbb{C}$: better to do so at the beginning of section 2.)}
% For the sake of brevity, we will omit the %subscription 
% {subscript}
% "$_\gPIC$" in the
% %following 
% {remaining}
% part of this section.

Within the MHD-gPIC model, we only %maintain 
record the guiding center position $\bm X_{k}$ rather than the full particle position $\bm x_k$. Consequently, we do not possess the momentum associated with the gyro-motion of the particles.
To address this limitation, we can derive from equation (\ref{eq:particle_location}) that
\begin{equation}
    \delta(\bm x-\bm x_k)\approx\delta(\bm x-\bm X_{k})-\delta\bm x_k\cdot\nabla\delta(\bm x-\bm X_{k}),\label{eq:delta_x}
\end{equation}
and by applying gyro-averaging to equation (\ref{eq:gPIC_momentum}), we obtain
\begin{equation}
    \bm F(\bm x,t)=\sum_k\dd{\bm P_{k}}t\delta(\bm x-\bm X_{k})-\sum_k\braket{\dd{\bm p_{k}}t\delta\bm x_k}\cdot\nabla\delta(\bm x-\bm X_{k}).
\end{equation}
Note that the term within the bracket $\braket{}$ is only a function of time, and hence can be placed behind the $\nabla$ operator.
By using $\D\braket{\bm p_k\delta\bm x_k}/\D t\sim\mathscr{O}(\omega_k^{-1})$ in the above equation, we can derive
\begin{equation}
    \bm F(\bm x,t)=\sum_k\dd{\bm P_{k}}t\delta(\bm x-\bm X_{k})+\nabla\cdot({\sf P}_{p\perp}(1-\bm{bb})),\label{eq:momentum_closure}
\end{equation}
where perpendicular pressure ${\sf P}_{p\perp}$ corresponds to gyro-energy $\gamma_km_k\omega_k^2r_k^2/2$.
%{[Should there also be a $\mu^2$?]}
%{[Briefly explain two things: dpk/dt=?, and how you manage to move nabla to the front, i.e., elaborate on the gyro-averaging process.]}
The first term on the right-hand side represents the momentum acquired by the particles, 
% while the second term is the perpendicular
% %-$\bm b$ 
% gradient of the stress tensor $\mathbf{T}=\sum_km_k\bm p_k\bm v_k\delta(\bm x-\bm x_{k})$.
% {(Double counting of $m_k$?)}
% For the first term, 
which is nearly equivalent to the
%electric force and 
Lorentz force acting on the particles. 
The difference lies in the fact that the charge density and current density are based on the guiding center position of the particles and include an additional gyro-averaged term. 
% The result is
% \begin{equation}
%     \sum_k\dd{\bm p_{gk}}t\delta(\bm x-\bm x_{gk})=
%     \rho_{e,\gPIC}\bm E-P_{\pe,\perp}\nabla\ln B+\sum_kq_k \bm v_{gk}\times\bm B,\label{eq:momentum_closure_1}
% \end{equation}
% where $\rho_{e,gPIC}$ is the charge density in gPIC part.
By employing equation (\ref{eq:electron_momentum_1}) and (\ref{eq:electron_momentum}), we can derive after some algebra (see Appendix \ref{appendix:eq_F} for details) 
\begin{equation}
\begin{aligned}
    \bm F=\rho_p\left.\DD{\bm u}t\right|_\perp+\nabla_\perp {\sf P}_{p\perp}+q_pE_\|\bm b+\Delta {\sf T}_p\bm \kappa+2\rho_p\Delta u_{p\|}\DD{\bm b}t,\label{eq:momentum_closure_2}
\end{aligned}
\end{equation}
where $\Delta {\sf T}_p= {\sf T}_{p\|}-{\sf P}_{p\perp}$ is the particles' stress tensor anisotropy in gas co-moving coordinates, and $\Delta u_{p\|}$ corresponds to particle bulk parallel velocity $u_{p\|}-u_\|$ in gas co-moving coordinates.

% where subscription $_\perp$ represents the perpendicular component, $\varepsilon\equiv\sum_k\gamma_km_k\delta(\bm x-\bm x_{gk})$ is the energy density 
%{(missing $\mathbb{c}^2$?)}%, $\rho_{q}$ is the charge density {(use $qn_p$?)}%, and $\Delta T= T_{\|}-P_\perp$ is the particles' stress tensor anisotropy in gas co-moving coordinate. 
%{(Particle stress tensor has not been defined yet... Also, it is getting more and more difficult to follow.)}

% If we have the momentum change of each particle, equation (\ref{eq:momentum_closure}) provides the closure of the macroscopic momentum change distribution.
% However, in the guiding center approximation, we know $\D p_{k\|}/\D t$ instead of $\D\bm p_{gk}/\D t$.
% By decomposing $\D\bm p_{gk}/\D t$ into its parallel and perpendicular components, we can express the summation of the momentum change as
% \begin{equation}
%     \bm F(\bm x,t)=
%     \sum_k\dd{p_{k\|}}t\delta(\bm x-\bm x_{gk})\bm b+T_{\|}\bm\kappa+p_\|\pp{\bm b}t.\label{eq:momentum_closure_3}
% \end{equation}
% Alternatively, we can predict the feedback from the macroscopic quantities of particles using equation (\ref{eq:momentum_closure_2}).

In a similar manner, by utilizing equation (\ref{eq:delta_x}) and applying the gyro-averaging technique, we arrive at
\begin{equation}
\begin{aligned}
    W(\bm x,t)&=\sum_k\dd{\varepsilon_{k}}t\delta(\bm x-\bm X_{k})-\sum_k\braket{\bm v_k\cdot\dd{\bm p_{k}}t\delta\bm x_k}\cdot\nabla\delta(\bm x-\bm X_{k})\\
    &=\sum_k{\dd{\varepsilon_{k}}t}\delta(\bm x-\bm X_{k})+\nabla\cdot\xkh{ {\sf P}_{p\perp} \bm u_{\perp}}\ .\label{eq:energy_closure}
\end{aligned}
\end{equation}
The second term on the right-hand side accounts for the energy flow resulting from the anisotropy of particle acceleration and de-acceleration during a single gyro-cycle.
%If we possess the energy gain data for each particle, equation (\ref{eq:energy_closure}) allows us to determine the change in energy density.
Given that particles acquire energy through acceleration by the electric field, by summing up all particles and applying gyro-averaging, we obtain
\begin{equation}
W=J_{p\|}E_\|+\xkh{2\rho_p\Delta u_{p\|}\DD{\bm b}t+\Delta \sfT_p\bm\kappa+\rho_p\DD{\bm u}t+\nabla {\sf P}_{p\perp}}\cdot\bm u_\perp,
\label{eq:energy_closure1}
% \begin{aligned}
%     \sum_k\dd{\varepsilon_k}t&\delta(\bm x-\bm x_{gk})=J_{\|}E_\| +(P_{\perp}\nabla_\perp\ln B+T_{\|}\bm\kappa)\cdot\bm v_{\text{MHD}}\\
%     &+P_{\perp}\p_t\ln B+\rho,\label{eq:energy_closure1}
% \end{aligned}
\end{equation}
where $J_{p\|}$ is the parallel current of the particle part, and the detailed {derivation} can be found in appendix \ref{appendix:eq_W}.
% where we denote the parallel current $J_{\|}\equiv \sum_k q_kv_{k\|}\delta(\bm x-\bm x_{gk})$.
% The three terms on the right-hand side correspond to acceleration by non-ideal electric field, acceleration by perpendicular electric field due to $\nabla B$-drift and curvature drift, and acceleration by induced electric field, respectively.
We can verify 
\begin{equation}
    W-\bm u\cdot \bm F=(J_{p\|}-q_p u_\|)E_\|\equiv Q,
\end{equation}
i.e., the energy gain in the gPIC part equals to work done on the MHD part plus the Joule heating.

\subsection{Equations of Gas Dynamics}\label{sec:gas_dynamics}
% {[Overall, it is extremely difficult to follow, even the logic itself. I suggest that you first outline the logic, and then fill in the underlying math. You can throw the relevant derivation to the appendix.]}
Built upon the particle dynamics {and feedback equations} described earlier, the MHD equations can be supplemented {with source terms from the gPIC component.}
%by enforcing conservation principles on local momentum and energy,
%Since the gPIC component does not impact the computation of the electromagnetic field, it is the momentum and energy equations that require modification. 
% We note that the gPIC-induced non-ideal electric field has already been described in Section \ref{ssec:Epara}, which modifies the induction equation. 
In this subsection, we focus on the momentum and energy equations. 
Naively, these gPIC source terms can simply be obtained by enforcing momentum and energy conservation, given by Equations (\ref{eq:momentum_closure_2}) and (\ref{eq:energy_closure1}). However, as we shall see in section \ref{sec:virtual_electron}, this approach is sensitive to inhomogeneous particle distribution especially Poisson noise, leading to large truncation errors.
By {adding gPIC perpendicular pressure to MHD part,}
% Although a subset of {particles} are modeled as particles, and the net charge of the remaining electrons is no longer zero, we introduce a set of virtual {particles}, which represent the anticipated influence of the gPIC part, to maintain symmetry in the equations. 
% {(update to include ions.)}
% By adding these virtual electrons to both sides of each equation, we preserve the left-hand sides of the MHD equations as they would be in a pure MHD system. 
%the particle feedback then functions as the source terms on the right-hand sides, 
we aim to retain the left-hand sides of equations (\ref{eq:momentum_feedback_predict}) and (\ref{eq:energy_feedback_predict}). As the MHD momentum equation is {coupled} with the equation of energy feedback (\ref{eq:energy_closure1}), {further derivation is needed to obtain} $\bm F_\gPIC$ and $W_\gPIC$.

\subsubsection{Source Term of Momentum Equation}
We begin our analysis with the momentum equation.
In our gPIC model, both electrons and ions are taken into consideration, and they
%have similar perpendicular-$\bm b$ velocities despite having 
{follow the same equation of motion yet have largely}
different masses.
%We neglect the bulk momentum of electrons, as in the pure MHD model, which results in different handling methods for ions and electrons.
{This requires us to handle particle electrons and ions differently, with particle electrons only supply pressure without bulk momentum.}
By summing the momentum equations for each component, we arrive at a comprehensive expression that captures the total momentum of the system
%{[I see here you deal with electrons and ions differently. This is something that should be noted before you do it.]}
\begin{equation}
    \rho_f\DD{\bm u}t+\sum_{k\in\text{ion}}\dd{\bm p_{k}}t\delta(\bm x-\bm x_k)+\nabla\cdot({\sf P}_f+\mathbf T_\pe)=\bm J\times\bm B,
\end{equation}
where the second term is a sum over the particle ions.
To further refine this equation, we employ the momentum closure equation (\ref{eq:momentum_closure_2}) specifically for ions. 

This allows us to rewrite the momentum equation as
% {[Should the terms on the RHS have negative sign according to Eq. 27?]}
\begin{equation}
\begin{aligned}
    \rho_f&\DD{\bm u}t+\left.\rho_p\DD{\bm u}t\right|_\perp+\nabla\cdot({\sf P}_f+\mathbf{T}_\pe)-\bm{J\times B}=\\
    &-\xkh{\nabla_\perp {\sf P}_{\text{pi}\perp}+q_\text{pi}E_\|\bm b+\Delta {\sf T}_\text{pi}\bm \kappa+2\rho_\text{pi}\Delta u_{\text{pi}\|}\DD{\bm b}t}.\label{eq:momentum_equation0}
\end{aligned}
\end{equation}
%{where $\Delta \sfT_p=\sfT_{p\|}-\sfP_{p\perp}$.}
In this formulation, we move the term $\nabla\cdot\mathbf{T}_\pe$ to the right-hand side and utilize the decomposition $\mathbf{T}_\pe = \sfT_{\pe\|}\bm{bb}+\sfP_{\pe\perp}(\mathbf{I}-\bm{bb})$. 

This transformation leads to a right-hand side that can be expressed as
\begin{equation}
    -\nabla\sfP_{p\perp}-\bm F_{p\perp}-\bm F_{p\|},\label{eq:rhs_momentum_feedback}
\end{equation}
where $\bm F_{p\perp}$ and $\bm F_{p\|}$ represent the perpendicular and parallel non-thermal forces, respectively, defined as
\begin{equation}
    \bm F_{{p}\perp}\equiv\Delta{\sf T}_{p}\bm\kappa+2\rho_\text{pi}\Delta u_{\text{pi}\|}\DD{\bm b}t,\label{eq:F_perp}
\end{equation}
and 
\begin{equation}
    \bm F_{p\|}\equiv\zkh{q_\text{pi}E_\|-\Delta {\sf T}_{\pe}\nabla_\|\ln B+\nabla_\|\xkh{{\sf T}_{\pe\|}-\sfP_{p\perp}}}\bm b.\label{eq:F_para}
\end{equation}
On the left-hand side, we are left with
\begin{equation}
    \rho_f\xkh{1+\frac{\rho_p}{\rho_f}(\mathbf{I}-\bm{bb})}\cdot\DD{\bm u}t+\nabla\sfP_f-\bm J\times\bm B.
\end{equation}
To simplify this expression, we introduce the ratio of the particle density to the total density,  $R\equiv\rho_p/(\rho_g+\rho_p)$, and express the inverse matrix of $1+\frac{\rho_p}{\rho_f}(\mathbf{I}-\bm{bb})$ as $1-R(\mathbf{I}-\bm{bb})$.
By moving $\nabla\sfP_{p\perp}$ to the left-hand side and multiplying $1-R(\mathbf{I}-\bm{bb})$ to both sides, we derive the final form of the momentum equation
% we can rearrange the momentum equation as following (the detailed derivation is given in appendix \ref{appendix:eq_F_gpic})
\begin{equation}
    \rho_f\DD{\bm u}t+(1-R(1-\bm{bb}))\cdot\xkh{\nabla\sfP-\bm J\times\bm B}=-(1-R)\bm F_{{p}\perp}-\bm F_{{p} \|}.
\label{eq:momentum_equation1}
\end{equation}
% where we defined 
%{[This is far from straightforward. Maybe you can include more derivations in the appendix. Also, I wonder where the $E_\parallel$ force is.]}
{This equation encapsulates the intricate interplay between the MHD fluid and the gPIC particles, highlighting the anisotropic properties of the gPIC component through $\Delta\sfT$ related terms in $\bm F_{p\perp}$ and $\bm F_{p\|}$.}
{If the gPIC component were in thermal and dynamical equilibrium with the MHD component, sharing the same temperature and bulk velocity, $F_{p\perp}$ would degenerate to zero as particles would drift with the MHD part together.}

{To maintain consistency with pure MHD format on the left-hand side of the momentum equation, we define the generalized gPIC force, $\bm F_{\gPIC}$ as}
\begin{equation}
    \bm F_{\gPIC}\equiv -R\xkh{\nabla\sfP-\bm J\times\bm B}_\perp+(1-R)\bm F_{p\perp}+\bm F_{p\|},\label{eq:F_gPIC}
\end{equation}
{which allows us to derive equation (\ref{eq:momentum_feedback_predict}), completing the formulation of the momentum equation within the gPIC framework.}

\subsubsection{Source Term of Energy Equation}
To derive the energy equation, we {first consider the energy exchange from the gPIC part.} Applying $\text{D}\bm u/\text{D} t$ from equation (\ref{eq:momentum_equation1}) 
\begin{equation*}
    \left.\DD{\bm u}t\right|_\perp=-\frac{1-R}{\rho_f}\xkh{\bm F_{p\perp}+\nabla_\perp\sfP-\bm J\times\bm B}
\end{equation*}
to equation (\ref{eq:energy_closure1}), leading to the following expression 
% {(see derivation in appendix \ref{appendix:eq_Wgpic})}%{[I don't get it.]}
\begin{equation}
\begin{aligned}
    W&=J_{p\|}E_\|+\xkh{\bm F_{p\perp}+\rho_p\DD{\bm u}t+\nabla \sfP_{p\perp}}\cdot\bm u_\perp \\
    &=J_{p\|}E_\|+\zkh{R(\bm J\times\bm B-\nabla\sfP)+(1-R)\bm F_{p\perp}+\nabla{\sfP}_{p\perp}}\cdot\bm u_\perp\\
    &=J_{p\|}E_\|+(\bm F_\gPIC+\nabla\sfP_{p\perp})\cdot\bm u_\perp.\label{eq:W0}
\end{aligned}
\end{equation}
%{[Is it possible to incorporate the thermal energy of virtual electrons to $\mathcal{E}$ so as to get rid of the $\nabla{\sf \tilde P}_\fe$ term?]}

{The other part of the energy equation involves the evolution of internal energy from fluid electrons and ions.}
{The ion part is more straightforward, where energy conservation is expressed as}
\begin{equation}
    \pp{\mathcal{E}_\text{fi}}t+\nabla\cdot\zkh{(\mathcal{E}_\text{fi}+\sfP_\text{fi})\bm u}=\bm E\cdot\bm J_\text{fi},
\end{equation}
{For fluid electrons, the equation in principle should be similar by replacing $\bm u$ by $\bm u+\Delta\bm u_\fe$. However, as massless fluid to neutralize the system, fluid electrons do not necessarily follow adiabatic evolution, and we also note that our single-fluid treatment implicitly assumes fluid electrons and ions share the same local temperature. Here we tentative adopt the following form}
%In contrast to the pure MHD case, the fluid electron velocity, $\bm u+\Delta \bm u_\fe$, may significantly differ from $u$ due to the inclusion of particle-represented electrons. %{[In this case, the induction equation also needs to be modified, i.e., the B field is advected at the speed of u+delta ufe.]}{[delta ufe only has parallel-b component]}
%{Taking this MHD electron velocity in the MHD energy equation, we note that the presence of the gPIC part makes it inappropriate to use an adiabatic closure for electrons in the gas since gPIC particles will not be heated by compression, which means the MHD electron bulk motion due to gPIC part (i.e., equation (\ref{eq:delta_u_fe})) only works for energy transmission.}
%{[Our underlying assumption about fluid electrons is that they share the same temperature as fluid ions. Does this hold?]}
%Therefore, we use the following form of fluid gas closure to reduce noise in the source terms
\begin{equation}
    \pp{\mathcal{E}_\fe}t+\nabla\cdot\zkh{(\mathcal{E}_\fe+\sfP_\fe)\bm u+\mathcal{E}_\fe\Delta \bm u_\fe}=\bm E\cdot\bm J_\fe.\label{eq:electron_energy_closure}
\end{equation}
The rationale behind this is that the baseline parallel motion of gPIC particles does not directly act on the bulk fluid. 
{Consequently, the redistribution of fluid electrons to balance the parallel current is treated as a combination of bulk motion (represented by the term above) and diffusion. The diffusion component can be incorporated, if necessary, together with the thermal conduction.}
%{[To reduce noise, this approach is based on intuition, the argument may not be sufficiently strong…]}{[I don't get why the last term in the bracket is multiplied by Efe, not Efe+Pfe.]}

%With the above MHD electron closure and MHD ion closure
%\begin{equation}
%    \pp{\mathcal{E}_\text{fi}}t+\nabla\cdot\zkh{(\mathcal{E}_\text{fi}+\sfP_\text{fi})\bm u}=\bm E\cdot\bm J_\text{fi},
%\end{equation}
Adding up the above, we obtain the MHD energy equation as
\begin{equation}
    \pp{}t\mathcal{E}_f+\nabla\cdot\zkh{\xkh{\mathcal{E}_f+\sfP}\bm u}-\bm{E}\cdot\bm J =\nabla\cdot\left(\sfP_{p\perp}\bm u-\mathcal{E}_\fe\Delta\bm u_\fe\right)-W,\label{eq:energy_equation3}
\end{equation}
where we used $W=\bm J_p\cdot\bm E$ and $\mathcal E_f=\mathcal E_\fe+\mathcal E_\text{fi}$.
Using equation (\ref{eq:W0}), the right-hand side can be rearranged into
\begin{equation}
    W_\gPIC = \bm F_\gPIC\cdot\bm u_\perp+J_{p\|}E_\| - \sfP_{p\perp}\nabla\cdot\bm u_\perp+ \nabla\cdot\xkh{\mathcal E_\fe\Delta\bm u_\fe-\sfP_{p\perp} u_\|\bm b},\label{eq:W_gPIC}
\end{equation}
where $\mathcal E_\fe$ can be calculated based on $\mathcal E_\fe/n_\fe=\mathcal E_f/n_f$, i.e., MHD electrons share the same temperature as fluid ions.

\subsection{Comparison with Previous Works}\label{sec:comparison}

We note that earlier studies have also investigated particle dynamics under the MHD framework using the guiding-center approach,
%\citep{Drake2019, Mignone2023}.
%In comparison with the backreaction-included model
in particular the \textit{kglobal} model \citep{Drake2019, Arnold2019, Yin2024}. Our MHD-gPIC model and implementation differ in several aspects including formulation, numerical method, and the base MHD framework.

First, our formulation preserves the Galilean invariance of particle motion — a property also achieved in \citet{Mignone2023} — by correcting $\Delta V_{k\|}$ rather than $V_{k\|}$ in Equation~(\ref{eq:electron_momentum}).
Additionally, we include the previously neglected energy feedback term, $\nabla\cdot\xkh{\sfP_{p\perp}\bm u_\perp}$, in equation (\ref{eq:energy_closure})\footnote{Here, we compare with the energy feedback formulation in \citet{Drake2019}, while \textit{kglobal} employs separate adiabatic equations for electrons and ions in the MHD component, as described in \citet{Yin2024}.}.
This term arise from the difference between the real positions of particles and its guiding centers.

Second, a key distinction in our numerical implementation 
%and \textit{kglobal} 
lies in the incorporation of $\sfP_{p\perp}$ into the MHD pressure for reasons explained earlier. In \textit{kglobal},
%Following their approach, 
this term is retained on the right-hand side (recall equation (\ref{eq:rhs_momentum_feedback})), and the MHD momentum and energy equations are formulated as
\begin{align}
    &\pp{\rho\bm u}t-\bm J\times\bm B+\nabla\cdot( \sfP_f+\rho\bm{uu})=-\bm F_\gPIC',\label{eq:momentum_feedback_appendix}\\
    &\pp{\mathcal{E}}t+\nabla\cdot\zkh{\xkh{\mathcal{E}+\sfP_f^*}\bm u+\frac c{4\pi}(\bm{E}-\bm E_0)\times\bm B}=-W_\gPIC',\label{eq:energy_feedback_appendix}
\end{align}
where $\sfP^*_f=\sfP_f+B^2/8\pi$. 
The terms $\bm F_\gPIC'$ and $W_\gPIC'$, representing particle feedback forces and work, are derived similarly to the methodology outlined in section \ref{sec:gas_dynamics},
\begin{align}
    &\bm F_\gPIC'=-R(\nabla P_f-\bm J\times\bm B)_\perp+\nabla \sfP_{p\perp}+(1-R)\bm F_{p\perp}+\bm F_{p\|},\\
    &W_\gPIC'=\bm F_\gPIC'\cdot\bm u_\perp+J_{p\|}E_\|+ \nabla\cdot\xkh{\mathcal E_\fe\Delta \bm u_\fe},
\end{align}
where the evolution of internal energy from fluid electrons is also taken into account.
The equation of state remains consistent with equation (\ref{eq:total_E}),  as in the pure MHD scenario.
In contrast to the above, our formulation is generally superior in noise mitigation, and offers distinct advantages when particle distribution is spatially inhomogeneous.
%in simulations where particle injection is localized or the particle distribution 
%is spatially inhomogeneous, since wave speeds in the Riemann solver are more accurately resolved.
The robustness of our approach under such conditions will be demonstrated through benchmark tests in section \ref{sec:virtual_electron}, and this is particularly important when considering more realistic particle injection prescriptions which likely occur in localized positions.

Third, the base MHD code in {\it kglobal} is very different from \verb|Athena++|. 
In particular, thermal electrons in {\it kglobal} evolve adiabatically and independently of the MHD fluid, which could lead to different electron and ion temperatures in the background fluid in an uncontrolled manner, where they are expected to share the same temperature in standard MHD as we use in \verb|Athena++|.
Furthermore, to mitigate numerical instabilities, \textit{kglobal} introduces hyper-diffusion for the MHD component and spatial diffusion for particles  \citep{Drake2019}.
These diffusive effects likely influence particle acceleration, as we discuss in section \ref{ssec:rec_compare}.

\section{Implementation of Guiding Center Method in Athena++}\label{sec:implementation}
{We have integrated the guiding center method into the \verb|Athena++| MHD code, building upon the MHD-PIC module described in \citet{Sun2023}.
Our implementation inherits the core infrastructure of the MHD-PIC framework—including particle storage within MeshBlocks, inter-block particle migration for adaptive mesh refinement, and field interpolation via the standard triangular-shaped cloud (TSC) scheme—without modification.
% The only notable extension to this inherited structure is the adoption of the van Leer time integrator for guiding center dynamics.
In this section, we focus on the novel components introduced specifically for the MHD-gPIC formulation and }
present the numerical methodology employed for particle integration and the procedural framework designed to handle particle feedback.

\subsection{Interpolation Variables}
Distinct from the MHD-PIC approach, the variables characterizing each particle in our guiding center method are the magnetic moment $\mu_k$ and the parallel momentum $P_{k\|}$ along the magnetic field direction, rather than the full three-dimensional momentum.
To determine the particle velocity $\bm V_k=V_{k\|}\bm b+\bm u_\perp$, which is essential for updating particle positions, we must interpolate the magnetic field direction $\bm b$ and the perpendicular gas velocity $\bm u_\perp$ based on the MHD framework for each particle.. 
In cases where higher-order perpendicular drift velocity $\bm v_\text{drift}$ is incorporated, additional computations are necessary for each particle, including $J_\|$, $\nabla B$, $\nabla_\|\bm u$ and $\bm \kappa$ (refer to equation (\ref{eq:v_drift})).
For the update of a particle’s parallel momentum, as dictated by equation (\ref{eq:particle_momentum}), the required variables are $\bm u$, $\nabla_\| B$, $\nabla_\|\bm u$ and $\bm \kappa$.
To streamline computational demands, we have identified a key relationship
\begin{equation}
    \nabla\cdot(\bm{bb})=\bm\kappa-\nabla_\|\ln B.
\end{equation}
This insight allows us to interpolate $\nabla\cdot\xkh{\bm{bb}}$ and directly compute the parallel and perpendicular components of $\bm b$, thereby eliminating the need for interpolating $\bm \kappa$ and $\nabla_\|B$.\footnote{The preference for  $\nabla\cdot(\bm{bb})$ over $\nabla\cdot(\bm{BB})$ is to circumvent numerical inaccuracies that arise when interpolated $B$ value is close to zero at the particle’s location.}
In essence, if the first-order perpendicular drift is neglected, we need to interpolate $13$ variables ($\bm u$, $\bm B$, $\nabla\cdot(\bm{bb})$, $E_\|$, $\nabla_\|\bm u$) for each particle at each step.
However, if higher-order terms are considered, the number of variables increases to $17$, necessitating the additional interpolation of $\nabla B$ and $J_\|$.
These cell-centered variables are essential for calculations that are performed at the cell centers. For clarity and consistency in our notation, we will use $[i,j,k]$ to denote the indices of the cells throughout the remainder of this document.

Numerically, the magnetic field $\bm B$ {can be accessed directly from the cell centers}.
To mitigate numerical errors, we employ the following approach for interpolating the gas velocity component
\begin{equation*}
    u_{x}[i,j,k]=\frac{(\rho_gu_x)[i-1/2,j,k]+(\rho_gu_x)[i+1/2,j,k]}{2\rho_{g}[i,j,k]},
\end{equation*}
where $(\rho_gu_x)$ represents the face-centered gas flux obtained from {Riemann solver}.
We prefer this approach over using cell-centered $u_x$ (similar for other components of $\bm u$) as we find it is more robust in handling of sharp velocity gradients.
%A similar approach is applied for interpolating other components of $\bm u$.
\footnote{For instance, in magnetic reconnection simulations, the gas velocity towards the center plane exhibits a sharp directional change near the center plane. Interpolating particle velocities from cell-centered $u$ can lead to particles settling towards the center plane, eventually decoupling from the gas mass distribution. This approach prevents such inaccuracies by providing a more accurate representation of the velocity field.}
For the calculation of spatial gradients, other variables necessitate a careful approach. 
Specifically, for $\nabla\cdot(\bm{bb})$, $\nabla B$ and $J_\|\equiv\bm b\cdot\nabla\times\bm B$, we rely on the face-centered magnetic field components. 
The face-centered $\bm B$ values are crucial for these calculations, as they provide a more accurate representation of the magnetic field variations within the computational cells.
For instance, in the $x$-direction, the face-centered $B_x$ can be directly accessed from the MHD component. 
However, for $B_y$ and $B_z$, we need to compute the averaged values from the nearby faces to ensure a proper interpolation. 
Specifically, the face-centered $B_y[i-1/2,j,k]$ at the cell interface in the $x$-direction is calculated as
\begin{equation*}
    \xkh{B_y[i-1,j,k]+B_y[i,j,k]}/{2}.
\end{equation*}
This averaging process is similarly applied to $B_z$ and in other directions, ensuring that both components are accurately represented at the cell interfaces.
For the remaining components of $E\|$ as derived from equation (\ref{eq:epara1})) and $\nabla_\|\bm u$, the calculations are performed directly using the cell-centered variables.

\subsection{Particle Integrator}
Based on these cell-centered variables, we can obtain all the necessary variables for updating particle positions and velocities. 
Within the primary computational loop, we employ the {two-stage} van Leer time integrator to push particles, achieving second-order temporal accuracy \citep{Stone2020}. In each time step, we use superscripts $^\ini$, $^\midd$, $^\fin$ to denote variables at the beginning, midpoint, and end of the step, respectively.
We optimize the particle integrator by performing the interpolation process only once per time step. 
This interpolation is based on the predicted midpoint positions of the particles, which effectively reduces the number of interpolation operations required.

In the $n$-th iteration loop, during the first stage, we determine the midpoint state by extrapolating from the results obtained in the preceding step. 
This process involves predicting the particles’ midpoint positions without updating their momenta. 
For particle $k$, the midpoint position $X_{k,n}^\midd$ is computed as follows
\begin{equation}
   \bm X_{k,n}^\midd = \bm X_{k,n}^\ini + \bm V^\ini_{k,n} \Delta t_n/2,
\end{equation}
where $\bm X_{k,n}^\ini\equiv\bm X_{k,n-1}^\fin$ denotes the initial position of the particle at the beginning of the time step, $\bm V_{k,n}^\ini$ is the initial velocity, and $\Delta t_n$ represents the time step.
The initial velocity $\bm V^\ini_{k,n}$ at $t=0$ is provided by the problem generator, and for subsequent steps, it is derived from the interpolated variables and the final velocity of the last step.
The magnetic field direction at the particle’s initial location can be approximated by
\begin{equation}
    \bm b_n^\ini\equiv\bm b^\fin_{n-1}\approx\bm b_{n-1}^\midd+\dd{\bm b^\midd_{n-1}}t\frac{\Delta t_{n-1}}{2},
\end{equation}
and the gas velocity is determined in a similar manner
\begin{equation}
\begin{aligned}
    &\bm u_{\perp,n}^\ini\equiv\bm u^\fin_{\perp,n-1}\approx\bm u_{\perp,n-1}^\midd+\dd{\bm u^\midd_{\perp,n-1}}t\frac{\Delta t_{n-1}}{2}\\
    &=\bm u_{\perp,n-1}^\midd+\dkh{\dd{\bm u^\midd_{n-1}}t\cdot\zkh{1-(\bm{bb})^\midd_{n-1}}-\bm u_{n-1}^\midd\cdot\dd{(\bm {bb})^\midd_{n-1}}t}\frac{\Delta t_{n-1}}2.
\end{aligned}
\end{equation}
We further assume $\D{\bm b}/\D t\approx \Delta V_{k\|}\bm\kappa$ and $\D{\bm u}/\D t\approx \Delta V_{k\|}\nabla_\|\bm u$ in the above equations, as $\text{D} \bm b/\text{D} t$ and $\text{D} \bm u/\text{D} t$ are unknown at particle's location. 
Combining these with
\begin{equation}
    \bm V_{k,n}^\ini=v_{k\|,n}^\ini\bm b_{n}^\ini+\bm u_{\perp,n}^\ini,
\end{equation}
and neglecting terms of $\mathscr{O}(\Delta t_{n-1}^2)$, we derive
\begin{equation}
\begin{aligned}
    \bm V^\ini_{k,n}=&\bm V^{*\fin}_{k,n-1}+\Delta {V^{\midd}_{k\|,n-1}}^2\bm \kappa^\midd_{n-1}\frac{\Delta t_{n-1}}2\\
    &+{\Delta V^{\midd}_{k\|,n-1}\xkh{\nabla_\|\bm u^\midd_{n-1}|_\perp-\bm u^\midd_{n-1}\cdot\bm\kappa^\midd_{n-1}\bm b}}\frac{\Delta t_{n-1}}2
    ,\label{eq:v_ini}
\end{aligned}
\end{equation}
where $\bm V^{*\fin}_{k,n-1}\equiv V^\fin_{k,n-1}\bm b^\midd_{n-1}+\bm u^\midd_{\perp,n-1}$ represents the particle velocity based on the final parallel momentum and midpoint interpolated variables from the previous step.
\footnote{Here, $V^\fin_{k,n-1}$ is based on the Lorentz factor calculated from $P_{k,n-1}^\fin$ and $\bm u_{\perp,n-1}^\midd$. 
This approximation is valid when $|u|\ll \C$.}
By incorporating the magnetic field line curvature, we enhance the precision of the prediction for $\bm V_{k,n}^\ini$ , under the assumption that the relative velocity between the particle and the gas does not undergo significant changes within a single time step.
{Right afterwards,} feedback from the gPIC component is calculated from equations (\ref{eq:momentum_feedback_predict}) and (\ref{eq:energy_feedback_predict}),
%To estimate particle feedback, 
{where} we must deposit gPIC quantities number density, parallel momentum, parallel and perpendicular pressure for each kind of particle.
{We note that these quantities are deposited based on particle locations at $t^{\rm mid}$ rather than $t^{\rm ini}$. This does not affect the order of accuracy for this stage (first order), while it allows us to recycle the particle weights in the next stage to reduce computational cost. }

The second stage of the time integration proceeds over a full time step, leveraging the midpoint state from the first stage as the foundation for the update. 
Utilizing the deposited particle number density and pressure from the initial stage, we determine the parallel electric field via equation (\ref{eq:epara1}). 
This provides the last necessary variables to evolve the particles’ parallel momentum using equation (\ref{eq:electron_momentum}).
We interpolate these required variables based on the cell-centered data obtained from the first part of this subsection at $t^\midd$, using particle $k$’s midpoint location $\bm X_{k,n}^\midd$. 
Subsequently, we calculate $\Delta P_{k\|}$ using equation (\ref{eq:electron_momentum}). 
With the parallel momentum $P^\midd_{\|k,n}=P^\ini_{\|k,n}+\Delta P_{k\|}/2$, equation (\ref{eq:v_gk}) yields the midpoint velocity $\bm V^\midd_{k,n}$. 
The particle position is then updated according to:
\begin{equation}
    \bm X^\fin_{k,n}=\bm X^\ini_{k,n} + \bm V^\midd_{k,n}\Delta t_n.
\end{equation}
Following this, $\bm V^\ini_{k,n+1}$ and $P^\ini_{k,n+1}=P^\ini_{\|k,n}+\Delta P_{k\|}$ are updated in preparation for the next iteration loop.
Finally, we deposit $\sfT_{p\|}$, $\sfP_{p\perp}$, parallel momentum $p_{p\|}$ for each particle species, and employ equations (\ref{eq:F_gPIC}) and (\ref{eq:W_gPIC}) to compute the feedback effects.

This two-stage integration process ensures a balance between computational efficiency and numerical accuracy, allowing for the precise simulation of particle dynamics within the evolving electromagnetic fields. 
The careful interpolation and update of variables at each stage are crucial for maintaining the consistency and stability of the simulation, particularly when dealing with complex interactions between particles and the surrounding plasma.

\subsection{MHD-gPIC Module in Athena++}
In this section, we provide a detailed outline of the numerical procedures that are specific to the gPIC component within the $n$-th iteration loop. 
Additionally, we introduce the numerical methods employed for calculating the feedback effects of the gPIC part on the overall system.
{It should also be noted that the  MHD-gPIC module, inherited from the MHD-PIC module, supports both static and adaptive mesh refinement, and we simply refer to \citet{Sun2023} for its implementation details.}

\subsubsection{Stage 1. Prediction}
In the first stage of van Leer time integrator, we aim to predict the midpoint state of the system.
\begin{itemize}
    \item \textbf{Stage 1.1: Particle Advancement} After computing hydro fluxes and integrating magnetic field, advance all particles by half a time step, $\Delta t_n/2$, using their initial velocities $\bm V^\ini_{k,n}$. This transition moves the particles from their initial positions $\bm X^\ini_{k,n}$ to their midpoint positions $\bm X^\midd_{k,n}$. These middle positions are the predict values, so their initial positions are still saved.
    \item \textbf{Stage 1.2: Weight Calculation} Calculate the particles’ weights over neighboring cells based on their midpoint positions  $\bm X^\midd_{k,n}$. This step is crucial for accurately distributing the particle properties onto the grid.
    \item \textbf{Stage 1.3: Deposition of Particle Properties} Deposit the parallel stress tensor $\sfT_{p,\|}$, perpendicular stress tensor $\sfP_{p,\perp}$, parallel momentum $p_{p,\|}$, and number density $n_{p}$ for each type of particle.
        From these deposited terms, compute the mass ratio $R=n_\text{pi}/(n_\text{pi}+n_\text{fi})$ and the parallel electric field $E_\|$ using equation (\ref{eq:epara1}).
        {Add the gPIC perpendicular} pressure $\sfP_{p\perp}$ to the MHD pressure before integrating the MHD part.
    \item \textbf{Stage 1.4: Particle Feedback Calculation} After integrating the MHD part, compute the particle feedback using equations (\ref{eq:F_gPIC}) and (\ref{eq:W_gPIC}).
        Specially, for the momentum feedback $\bm F_\gPIC$, we notice that
        \begin{equation*}
            -\nabla\cdot(\rho_g\bm{uu}+\sfP)+\bm J\times\bm B=\pp{\rho_g\bm u}t
        \end{equation*}
        has already been calculated in the Riemann solver of the MHD part.
        Therefore, the momentum feedback is calculated by
        \begin{equation}
            \bm F_\gPIC=R\xkh{\pp{\rho_g\bm u}t+\nabla\cdot(\rho_g\bm{uu})}_\perp+(1-R)\bm F_{p\perp}+\bm F_{p\|},
        \end{equation}
        where $\nabla\cdot(\bm{bb})$, $\nabla_\|\bm u$ and $E_\|$ are calculated here to compute $\bm F_{p\perp}$ and $\bm F_{p\|}$ using equations (\ref{eq:F_perp}) and (\ref{eq:F_para}).
        \footnote{If high-order perpendicular drift of particle is included in simulation, $\nabla B$ and $J_\|$ are also calculated here.}
        Add the momentum and energy source terms from the particles based on equation (\ref{eq:F_gPIC}) and (\ref{eq:W_gPIC}) to the gas as source terms.
        To avoid numerical problems, we apply the following limiter for  $\Delta\bm u_\fe$ when computing energy feedback:
        \begin{equation}
            \Delta \bm u_\fe=\xkh{(n_\pe-n_{\rm pi}) u_\|\bm b+\frac{J_{p\|}\bm b}{e}}\Delta V/\max\xkh{n_\fe\Delta V,\sqrt{n_p\Delta V}},
        \end{equation}
        where $\Delta V$ represents the cell volume in the simulation. 
        {Since the calculation of $n_{\fe}\Delta V$ is subject to an error on the order of $\sqrt{{n}_{p}\Delta V}$ due to particle number fluctuations in the gPIC module, the computed $n_{\fe}\Delta V$ can become zero in regions where particle electrons dominate, leading to numerical instability. To prevent this, the limiter imposes a lower bound on $n_{\fe}\Delta V$ equal to the magnitude of this fluctuation.}
        Although the problem setup should be designed to prevent such situations by employing a sufficiently large number of particles to accurately represent a specific portion of the plasma, the inclusion of this limiter serves as a safeguard against extreme scenarios that could arise due to numerical fluctuations or insufficient particle sampling.
    \item \textbf{Stage 1.5: Application of Boundary Conditions} Apply boundary conditions to both the MHD gas and particles. 
\end{itemize}

In this stage, the deposition occurs after particles have moved half a step, allowing for the calculation of the particle number density and stress tensor at the midpoint. 
This midpoint evaluation is also crucial for the accurate calculation of non-ideal electric fields, as these fields are subsequently used to update particle momentum during the second stage of the simulation.
Furthermore, since the feedback mechanism in the second stage also relies on the particle distribution at the midpoint, we can optimize the process by avoiding the redundant deposition of the particle number density. 
%This optimization not only streamlines the simulation but also conserves computational resources.
%Regarding particle feedback, it is noteworthy that the choice between using the initial or midpoint particle state does not compromise the precision of the simulation. 
%The feedback is designed to be robust regardless of this choice.

\subsubsection{Stage 2. Advance}
In the second stage, the entire system is advanced from the initial state to the final state, leveraging the predicted midpoint state as a basis for this progression. 

\begin{itemize}
    \item \textbf{Stage 2.1: Re-Compute $E_\|$} In stage 1.4, the parallel electric field is calculated based on particle pressure at midpoint time and gas pressure at initial time. To ensure precision, we compute $E_\|$ once again based on midpoint gas pressure for updating the particle's momentum.
    \item \textbf{Stage 2.2: Interpolation} For each particle $k$, interpolate $\bm{u}$, $\bm{B}$, $\nabla\cdot(\bm{bb})$, $E_{\|}$, $\nabla_{\|}\bm{u}$, $\nabla B$ and $J_{\|}$ (when high-order perpendicular drift is considered). Compute the local magnetic curvature and the parallel magnetic gradient based on $\nabla\cdot(\bm{bb})$.
    \item \textbf{Stage 2.3: Advance Particles} Compute $\Delta P_{k,n}$ with equation (\ref{eq:electron_momentum}) based on the interpolated variables and update the particle's parallel momentum from $P^\ini_{k,n}$ to $P^\midd_{k,n}$. Find particle velocity $\bm{V}^\midd_{k,n}$ by equation (\ref{eq:v_gk}) and update the particle's position $\bm{X}^\fin_k$ based on $\bm{X}^\ini_k$. Calculate $\bm{V}^\ini_{k,n+1}$ with equation (\ref{eq:v_ini}) for the next loop. Update the particle momentum to $P^\fin_{k,n}$.
    \item \textbf{Stage 2.4: Deposition of Particle Properties} Deposit $T_{p,\|}$, $P_{p,\perp}$, $p_{p,\|}$ for each kind of particle.
    \item \textbf{Stage 2.5: Particle Feedback Calculation} Integrate the MHD gas from $t^\ini$ to $t^\fin$, and then subtract the particle momentum and energy gain computed with equations (\ref{eq:momentum_equation1}) and (\ref{eq:W_gPIC}) similar to stage 1.4. Note that the computation of $\nabla\cdot(\bm{bb})$ is not necessary at this stage, as it has already been calculated using the midpoint magnetic field during the first stage. 
    \item \textbf{Stage 2.6: Application of Boundary Conditions} Apply appropriate boundary conditions to both the MHD gas and particles.
\end{itemize}

In this stage, the feedback calculation is based on the particle state at midpoint time, considering both position and momentum. 
This ensures that the entire algorithm is in alignment with the two-stage van Leer integrator. 
However, there is a caveat: the non-ideal electric field calculation relies on the particle momentum at the initial time, but at the midpoint location.
Given that this electric field is interdependent with particle momentum, achieving the same precision as the two-stage van Leer integrator would necessitate an additional round of deposition and interpolation. 
Nevertheless,
%expending further computational resources on this refinement is not justified. This is because the current parallel electric field is an approximation and should be improved in the future.
{we refrain from pursuing such additional refinement since the parallel electric field itself is already approximated.}

\subsubsection{Riemann Solver}

Our MHD equations use modified pressure:
%Note that the equations solved in the Riemann solver differ from the standard MHD equations: 
$\mathsf{P}$ is used which includes contribution from particles $\mathsf{P}_{p\perp}$ instead of $\mathsf{P}_f$.
%This will modify the wave speeds and hence the Riemann solver itself should be modified as well. 
{On the other hand, the definition of energy density uses $\mathsf{P}_f$ instead of $\mathsf{P}$, thus the system is not equivalent to the MHD equations in standard form.}
%The particle pressure $\mathsf{P}_{p}$ couples with the total pressure and influences the wave speeds, so we include $\mathsf{P}_{p\perp}$ to represent this effect.
%Although pressure anisotropy in the gPIC component would complicate the wave speeds, we account for this effect as a feedback mechanism.

%This approach modifies the original derivation of the Riemann solver. 
%To justify it, consider the following:
%\begin{itemize}
%    \item The gradient of $\mathsf{P}_{p\perp}$ and the divergence of $\mathsf{P}_{p\perp}\bm{u}$ are computed in the same manner as their counterparts in the standard Riemann solver.
%    \item For convenience, we move these terms to the left-hand side.
%\end{itemize}
The wave speeds in the Riemann solver are based on $\mathsf{P}$ rather than $\mathsf{P}_f$ because this provides more accurate results---particularly for isotropic particles and the electron gPIC case.
The advantage of this treatment is demonstrated in Section~\ref{sec:virtual_electron}, where severe noise arises without it.

\begin{figure}
    \centering
    \includegraphics[height=5.7cm, width=8.5cm]{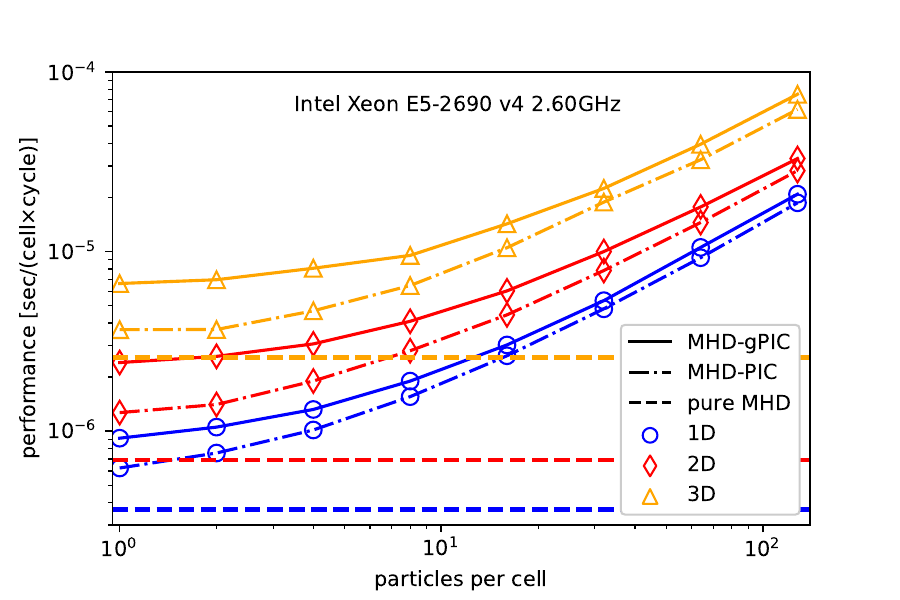}  
    \caption{The performance for MHD-gPIC compare to pure MHD and MHD-PIC, with the test in Section \ref{sec:cpaw}. The total time cost for integrating one cell relies on the number of particle per cell, as the sum of pure MHD cost and particle cost. The precision is in double and the simulation dimensions are referred to colors.}
    \label{fig:benchmark}
\end{figure}

\subsection{Performance Test}

%To evaluate the computational efficiency of our code, 
To assess the performance our MHD-gPIC module, we employ the circularly polarized Alfvén wave test (Section \ref{sec:cpaw}) with varying particle number per cell in one to three dimensions.
%, utilizing double precision for our calculations. 
The code is compiled using the GNU compiler and executed on a single compute node equipped with two Intel Xeon E5-2690 v4 CPUs, providing a total of 28 cores. {As the handling of MPI communication is identical as in \citet{Sun2023}, we do not repeat the scaling test there.}
%We conducted a series of tests, varying the number of particles per cell and performing simulations in one, two, and three dimensions to ascertain the computational cost, which we 
Code performance is quantified as the time (in seconds) required for each cell update per time-step.
The findings from our MHD-gPIC, MHD-PIC, and pure MHD simulations are depicted in Figure \ref{fig:benchmark}, with distinct line styles and colors used to differentiate between the methods. 
%It is worth noting that 
Note that in the MHD-gPIC simulations, high-order perpendicular drift of particles is not incorporated. 
By establishing the pure MHD simulation performance as a baseline, we are able to dissect the computational demands imposed by particle dynamics.

In scenarios with a low particle count, the gPIC module introduces a more significant computational overhead compared to the MHD-PIC module, approximately 70\% of the MHD cost. 
This is attributed to the necessity of calculating additional interpolated variables and the computation of momentum-energy feedback from the particle component based on deposited variables. 
In contrast, the MHD-PIC module only requires interpolation of $\bm u$ and $\bm B$, with direct use of deposited variables for feedback.
As the particle number increases,
%the updating of individual particle states becomes the primary computational burden. 
{individual particle updates take over to dominate the computational cost.}

Our simulation data indicates that the additional computational cost per particle {relative to MHD-PIC} is approximately 10\% for 1D simulations, 13\% for 2D simulations, and 18\% for 3D simulations. 
Despite interpolating 13 variables per particle in the MHD-gPIC module (or 17 variables when including high-order perpendicular drift) during the second stage—compared to the 6 variables in the MHD-PIC module—the performance remains well within acceptable limits. This is particularly encouraging when considering the MHD-gPIC’s advantage of permitting significantly larger numerical time steps. We further rerun the tests including high-order perpendicular drift, which we find to increases the cost per particle by approximately 6\% for each dimension compared to simulations without high-order perpendicular drift. The marginal increase in cost, combined with the enhanced accuracy and stability provided by the high-order terms, underscores the cost-effectiveness of our MHD-gPIC approach.

\section{Code Tests}\label{sec:tests}
{In this section}, we assess the accuracy and efficacy of our MHD-gPIC implementation by presenting a suite of benchmark tests in this section.

\subsection{Trajectory Tests}\label{sec:loop}
\begin{figure*}
    \centering
    \includegraphics[height=9.6cm, width=18cm]{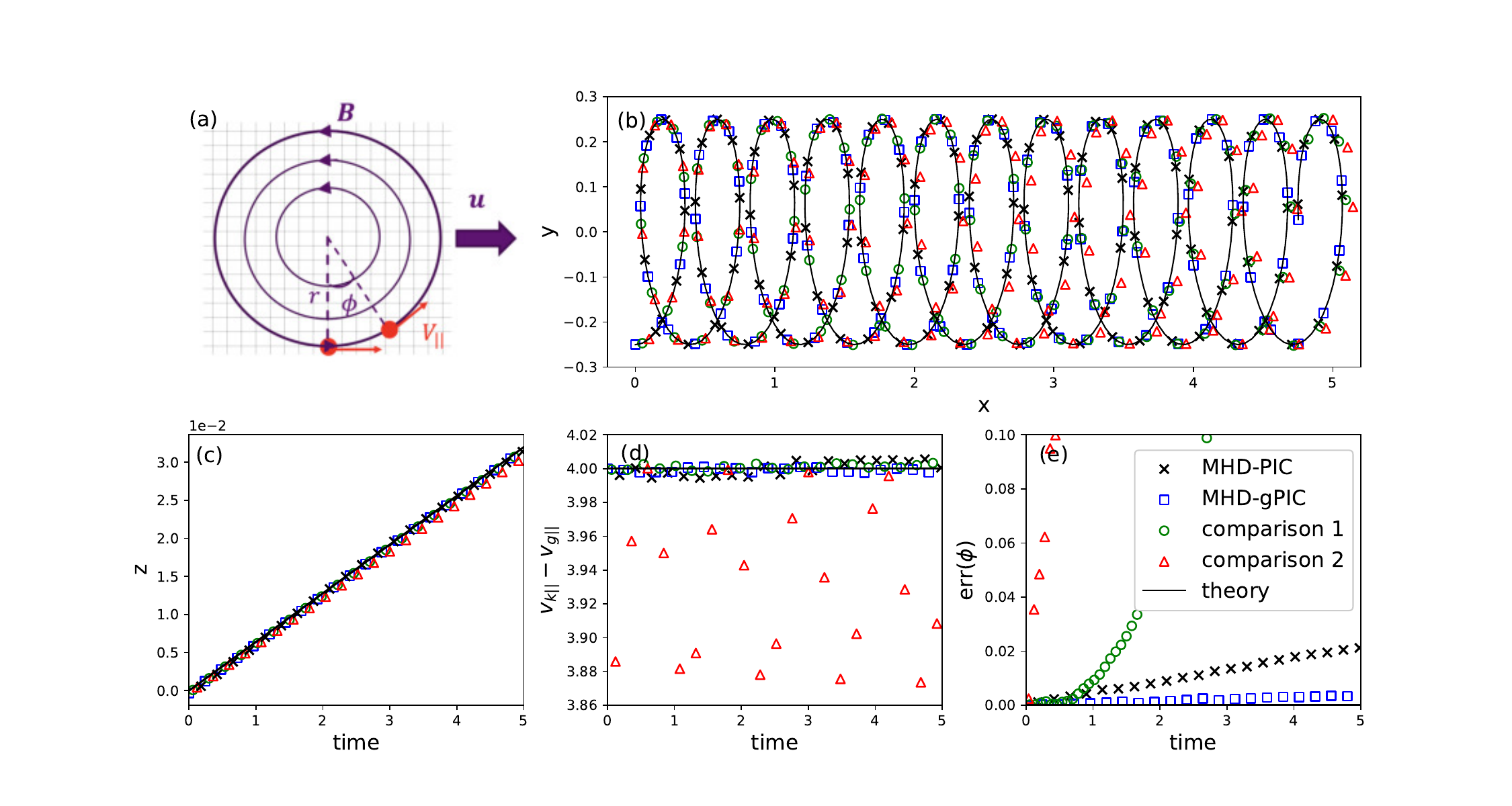}  
    \caption{Numerical result of a particle moving with a 2D magnetic field loop. (a) Schematic diagram of the test. (b) The trajectory in x-y plane, (c) drift in z direction, where shadowed region {(though not quite visible)} stands for $z\pm r_{L}$. (d) Parallel velocity in gas co-moving coordinate. (e) Phase error. Different colors/symbols indicate simulations with the MHD-PIC module, MHD-gPIC module and its variants, labeled in the legend.}
    \label{fig:drift}
\end{figure*}

To evaluate the precision of the particle integrator, we established a 2D magnetic field loop configuration as described in \citep{Gardiner2005} with a hot particle without feedback, as depicted in Fig. \ref{fig:drift} (a).
The ambient gas was set in motion with a velocity of $\bm u=u_0\hat x$, and the magnetic field was initialized with a vector potential of $\bm A=A_0\max(R-r,0)\hat z$.
The particle was injected at a position of $-r\hat y$ with an initial parallel velocity of $v_0$.
The light speed was chosen to be $\mathbb C\gg v_0, u_0$, ensuring that the particle's motion is non-relativistic.

The theoretically predicted trajectory of the hot particle is anticipated to trace the magnetic field lines within the $x-y$ plane. 
The particle's trajectory, as observed in the simulation frame, can be expressed by the following equations:
\begin{equation}
\begin{aligned}
    x_0(t)&=u_0t+r\sin\phi,\\
    y_0(t)&=-r\cos \phi,\\
    z_0(t)&=v_{\bm{\kappa}}t,
\end{aligned}\label{eq:x0}
\end{equation}
where $\phi(t)$ represents the phase in the co-moving frame. 
In this co-moving frame, where the gas is at rest, the particle should follow the field line within the $x-y$ plane, implying $\phi=(v_0-u_0)t/r$.
While curvature drift and gradient drift are often considered negligible within the guiding center approximation,
{they are incorporated in our implementation.}
Specifically in this test, $v_{\bm\kappa}=(v_0-u_0)^2/r\Omega_k$ {is computed automatically, but we manually set} $v_{\nabla B}=0$ {to avoid sharp gradient at loop edges (for testing purposes)}.

The magnetic field loop has a radius of $R=0.4$ with an initial vector potential $A_0=0.001$, and the ambient gas velocity is set at $u_0=1.0$.
We inject a particle whose initial position is $[0,-0.25]$ with an initial parallel velocity of $v_0=5.0$. Its magnetic moment is set by a gyro-radius of $r_L=3\times10^{-4}$, which corresponds to approximately 1\% of the grid size.
To provide a comparison baseline, we {also} conduct the same test using the MHD-PIC module. 
{In this case}, the MHD-PIC module's time step is approximately $160$ times {smaller} than that of the MHD-gPIC module due to the imposed constraint of the {gyration} CFL number, which limits the time step.
Additionally, we perform two comparison tests to further evaluate the impact of different numerical approaches. 
In the first comparison test, we omit the correction term in equation (\ref{eq:v_ini}).
In the second comparison test, we use the approximation $\D \bm b/\D t=\p \bm b/\p t$ based on the original formula presented by \citet{northrop1963} (see equation (\ref{eq:electron_momentum_0})).
We label these tests as “comparison 1” and “comparison 2” for ease of reference. 

Figure \ref{fig:drift} (b) and (c) displays the simulation trajectories. 
Both the MHD-PIC module and the MHD-gPIC module yield trajectories that align closely with the theoretical predictions, while the results from the comparison tests deviate significantly. 
This deviation is attributed to the inability to maintain a constant relative velocity between the particle and the gas in the comparison tests.
Panel (d) illustrates the evolution of parallel velocities in the gas co-moving coordinate system. 
The particle velocity from the MHD-gPIC test remains nearly constant, with the error in this test arising from inaccuracies in the computation of magnetic curvature. 
For comparison 1, the relative velocity oscillates slightly due to the direct use of the midpoint velocity from the last step during the simulation. 
A smaller time step would be required to achieve results comparable to those with the correction term. 
As for comparison 2, the error is due to the negligence of the differences between $\D \bm b/\D t$ and $\p \bm b/\p t$ in the particle’s momentum equation. This difference is only negligible when particle velocity is similar to gas velocity and which is hardly applicable for energetic particles.
By comparing the phase angle error of particles within the magnetic loop (depicted in panel (e)), it is evident that, despite the significantly larger time step of the MHD-gPIC approach compared to the MHD-PIC method, the former exhibit {even smaller error}. 
This demonstrates the robustness of the MHD-gPIC module in handling larger time steps without compromising accuracy.

Through this test, we have validated that the MHD-gPIC module can accurately integrate particle trajectories without being constrained by the gyration CFL number. 
The use of equation (\ref{eq:dbdt}) helps to avoid errors in the integration of particle momentum, thereby allowing us to significantly increase the minimum time step required for particle simulations. 
This advancement has the potential to enhance the efficiency of particle-in-cell simulations, making them more feasible for large-scale and long-duration studies.

\subsection{Circularly Polarized Alfvén Wave}\label{sec:cpaw}
To verify the precision of the feedback from energetic particles on plasma dynamics, we carry out a series of numerical tests employing circularly polarized Alfvén waves (CPAW) {that are modified by the presence of energetic particles, including the onset of} the firehose instability \citep{Drake2019}.
The MHD component is configured with a standard CPAW setup, featuring a constant background magnetic field with a minor perpendicular perturbation. 
The presence of particles cab induce a pressure anisotropy within the plasma, which influences the Alfvén wave dispersion relation. 
In a uniform magnetic field of strength $\bm B_0$, when the total parallel pressure $P_\|$ and the total perpendicular pressure $P_\perp$ of the plasma are unequal, the momentum equation becomes
\begin{equation}
    \rho\dd{\bm u}{t}=-\nabla\left(\sfP_\perp+\frac{B_0^2}{8\pi}\right)+\nabla\cdot\left[\bm{bb}\left(\sfP_\perp-\sfP_\|+\frac{B_0^2}{4\pi}\right)\right].
\end{equation}
Considering perturbations that are perpendicular to the background field, the dispersion relation is given by
\begin{equation}
    \omega^2\rho=k^2\left(\sfP_\perp-\sfP_\|+\frac{B_0^2}{4\pi}\right),\label{eq:cpaw}
\end{equation}
where $\rho$ represents the background plasma density. 
If $\sfP_\perp-\sfP_\|+B_0^2/{4\pi}>0$, one obtain a modified CPAW wave,
%velocity can be predicted based on the dispersion relation, 
while the wave will grow exponentially if $\sfP_\perp-\sfP_\|+B_0^2/{4\pi}<0$, a condition known as the firehose instability.
{The resulting eigen-states under a uniform background consist of perpendicular magnetic field in circularly polarized configuration, with perpendicular velocity given by}
%The initial gas velocity and magnetic field perturbation follows the relationship:
\begin{equation}
    \delta \bm u_\perp = -\frac{\omega}{k}\delta \bm B.
\end{equation}
{Note that the system is degenerate between left and right polarizations.}
%{(Please note that I have rearrange your text here. State the eigenvectors right after the dispersion relation, and then you can focus on problem setup. Also, please pay attention to the use of $4\pi$ in physical derivations vs in the code. You may consider making a statement somewhere.)}
 
In our tests, the direction of the background magnetic field and the wave’s propagation direction are aligned along the $x$-axis in 1D simulations, and are oriented diagonally with respect to the grid in 2D and 3D scenarios. 
The gPIC component is initialized with pressure anisotropy, characterized by different temperatures for the parallel momentum and magnetic moment. If the parallel pressure of electrons in the gPIC part exceeds their perpendicular pressure, the perpendicular temperature of the particle electrons is set to match that of the background gas, and the converse is true if the perpendicular pressure is greater. 
The temperature of ions in the gPIC component is consistently initialized to be equivalent to that of the background gas.
Particles’ velocities are set up based on a Maxwell distribution with $v_{th,i}=c_s/\sqrt{2}$ for ions and $v_{th,e\perp}=\sqrt{m_i/m_e}v_{th,i}$, $v_{th,e\perp}=\sqrt{m_i\sfP_{\pe\|}/m_e\sfP_{\pe\perp}}v_{th,i}$ for electrons, where $c_s$ is the sound speed.
The mass ratio $m_i/m_e$ between ions and electrons is set at $25$ in this test.
The computational units are defined as $\rho=B_0=2\pi/k=1$. 
The simulation box size and resolution, specified as $(L_x/N_x, L_y/N_y, L_z/N_z)$, are (1/256) for 1D, (1/256, 0.25/64) for 2D, and (1/256, 0.25/64, 0.25/64) for 3D.
Eigen-state perturbations are initialized with an amplitude of $\delta B=10^{-2}B_0$.
We employ $100$ electrons per cell to represent $20\%$ of the total electrons in each test. 
Additionally, we introduce 100 ions per cell in a separate set of tests, which corresponds to $50\%$ of the total ion population for tests with particle ions.
\begin{figure}
    \centering
    \includegraphics[height=8 cm, width=8cm]{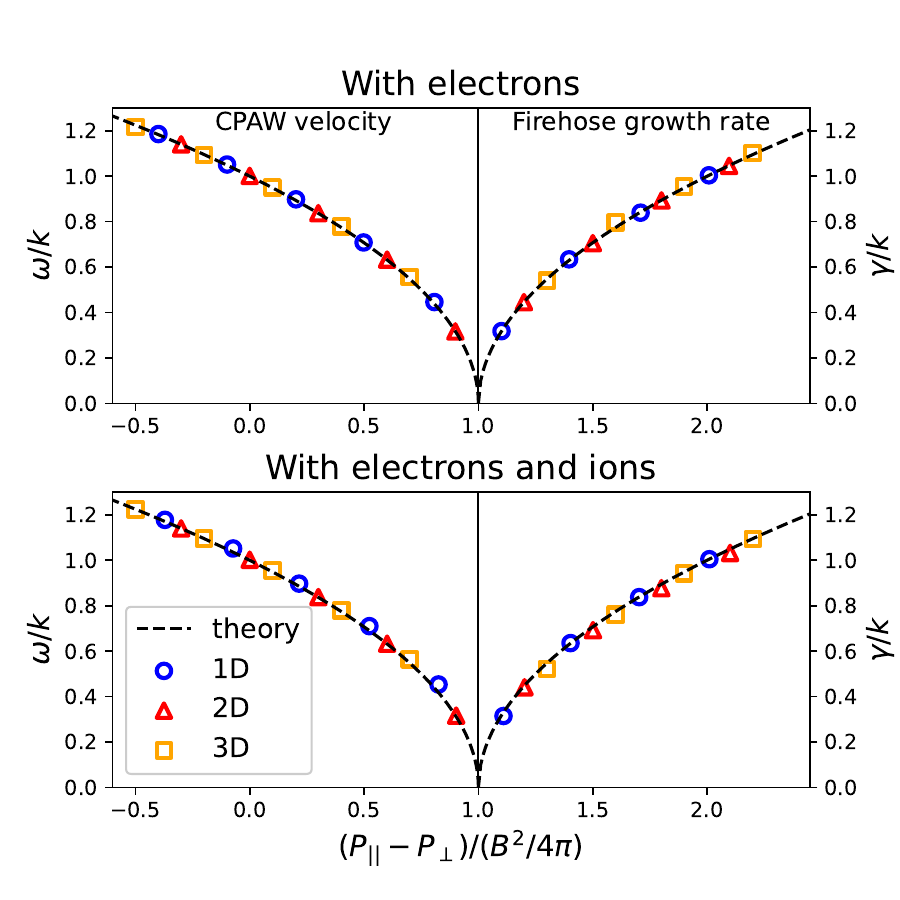}  
    \caption{
    CPAW and firehose instability test results. The upper panel shows simulation results of tests with particle electrons, and the lower panel shows simulation results of tests with both electrons and ions in the gPIC part. The left half of both panels shows the phase speed of modified CPAW, and the right half shows the growth rate of the firehose instability, both as a function of pressure anisotropy. Simulation results from different dimensions are plotted against analytical theory.
    % CPAW and firehose instability test results. The upper panel shows simulation results of tests with particle electrons and the lower panel shows simulations results of tests with both electrons and ions in gPIC part. The left half of both panels shows {the phase speed of modified} CPAW, and the right half shows the growth rate {of the firehose instability, both are shown} as a function of pressure anisotropy. Simulation results from different dimensions are plotted agains analytical theory.
    }
    \label{fig:firehose}
\end{figure}

By adjusting the parallel and perpendicular pressures of the particle electrons, we can analyze the phase speed of the CPAW under various pressure anisotropy conditions and compare these results to the analytical predictions. 
Equation (\ref{eq:cpaw}) yields the phase velocity of the modified CPAW as
\begin{equation}
v=\sqrt{1+\frac{\sfP_\perp-\sfP_\|}{\rho V_A^2}}V_A,
\end{equation}
where $V_A$ denotes the Alfvén velocity of the background gas. The theoretical and simulation results across different dimensions and with or without ions are presented in the left half of two panels of Figure \ref{fig:firehose}, which show excellent agreement.

Once $\sfP_{||}-\sfP_\perp$ exceeds $\rho V_A^2$, it marks the onset of the firehose instability, with the growth rate from Equation (\ref{eq:cpaw}) given by
\begin{equation}
    \gamma = \sqrt{\frac{\sfP_\|-\sfP_\perp}{\rho V_A^2}-1}V_Ak.
\end{equation}
The growth rate is proportional to the wave number $k$, indicating a strong dependence on the spatial scale of the perturbations.
To quantitatively assess this relationship, we measure the growth rate of the $2\pi/k=1$ mode across various pressure anisotropy configurations. 
The simulation data, presented in the right half of the panels in Figure \ref{fig:firehose}, exhibit a remarkable agreement with the analytical solution.
%thereby validating our numerical approach and its capability to accurately capture the dynamics of the firehose instability. 
%This consistency between theory and simulation underscores the reliability of our model for investigating plasma instabilities and their subsequent effects on wave propagation and particle behavior in anisotropic plasmas.
{This test demonstrate the accuracy in the treatment of particle momentum feedback from our MHD-gPIC model.}

\subsection{Electron Acoustic Wave}\label{sec:eaw}

\begin{figure}
    \centering
    \includegraphics[height=3.94 cm, width=9cm]{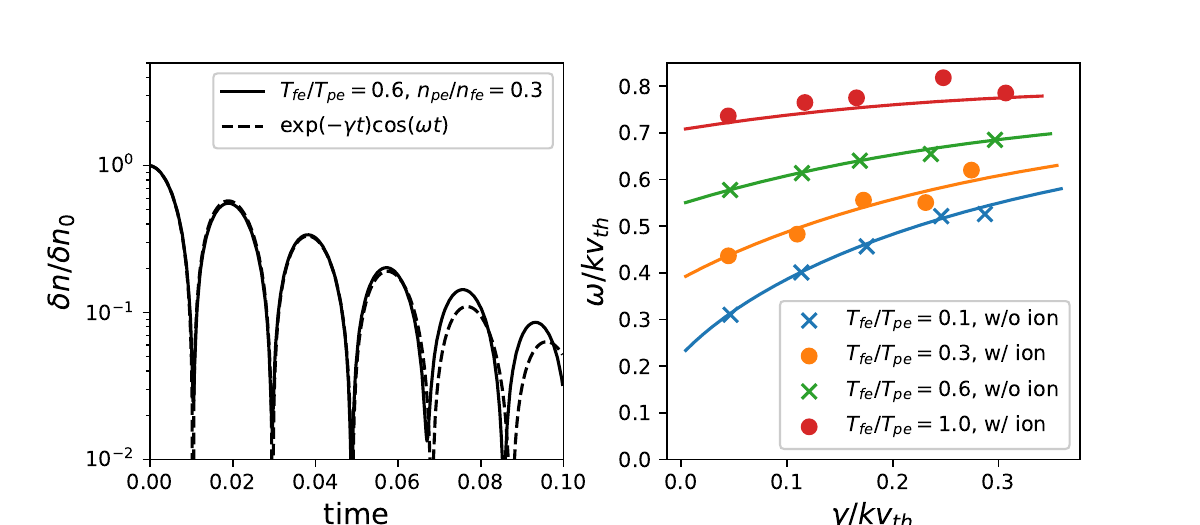}  
    \caption{EAW angular frequency as a function of damping rate. Different colors corresponds to different $n_c/n_h$, lines are theoretic solution. Dots of same color are from simulations with different $T_c/T_h$.}
    \label{fig:eaw}
\end{figure}

%We next aim to verify the precision of the MHD-gPIC method 
{In this subsection, we aim to further}
test our formulation and implementation of parallel electric fields by simulating Landau damping of electron acoustic waves (EAW) as in \citet{Arnold2019}.
In our setup, the gas and particles are characterized by distinct temperatures. The particles are initially distributed according to the Maxwell distribution with a density profile given by
\begin{equation}
    n(x) = n_0+\delta n\cos(2\pi x/L),
\end{equation}
where $\delta n$ denotes the amplitude of the perturbation, and $L$ is the length of the simulation domain. 
The feedback from the hot particles is considered negligible due to the smallness of $\delta\rho$, resulting in a nearly constant gas density throughout the simulation.
{From equation (\ref{eq:epara1}), the parallel electric field for particles with isotropic pressure is determined by}
\begin{equation}
    E_\| = -\frac1{n_ie}\left(\nabla P_\pe-\frac{\nabla n_\pe}{2n_\text{fi}} P\right).
\end{equation}
The random motion of the particles leads to a decay in the perturbation amplitude over time, with $\gamma$ representing the damping rate in this context.
Since the perturbation in our setup is a standing wave, the presence of the electric field modifies its angular frequency $\omega$, a phenomenon known as Landau damping.
The dispersion function for EAW is
\begin{equation}\label{eq:eaw}
    \frac{n_\text{fe}}{n_\text{pe}}=Z'(\zeta)\left(\frac{T_\text{fe}}{2T_\pe}-\zeta^2\right),
\end{equation}
where $\zeta = (\omega-i\gamma)/kv_\text{th}$, $v_\text{th}$ is the particle thermal velocity and $Z(\zeta)$ is the plasma dispersion function.
The expected evolution of the particle density is
\begin{equation}
    n(x,t)=n_0+\delta n\cos(2\pi x/L)\cos(\omega t)e^{-\gamma t}.
\end{equation}

We conduct 1D simulations with $64$ uniform grids and total $1.024\times10^6$ electrons in the gPIC part. 
These electrons are initialized by a Maxwell distribution with different temperatures from the MHD gas.
The gas density is uniform.
We also add ions to the gPIC composition in part of the tests. The ion number density is 
$1.6\times10^4$ per cell, which represents $20\%$ of the total ions. 
They have the same temperature as the MHD gas.
We tested the wave frequency and damping rate across varying ratios of $n_\text{fe}/n_\pe$ and different values of $T_\text{fe}/T_\pe$.
In the left panel of Figure \ref{fig:eaw}, we plot the evolution of gPIC electron perturbation amplitude with $n_\text{fe}/n_\pe=0.6$ and $T_\text{fe}/T_\pe=0.3$ against the theoretical result. 
The simulation results reveal excellent agreement with theoretical results over multiple periods.
Then we fit the values of $\omega$ and $\gamma$ for each test using the time-evolution of the perturbation amplitude.
The results are depicted in the right panel of Figure \ref{fig:eaw}, where it is evident that our simulation outcomes align well with the analytical predictions.

\subsection{The Role of $\sfP_{p\perp}$ in $\sfP$ for Noise Reduction}\label{sec:virtual_electron}

We derived the MHD momentum and energy equations with {$\sfP_{p\perp}$ in $\sfP$} in section \ref{sec:gas_dynamics} and list these equations without it in section \ref{sec:comparison}. 
In this section, we discuss the necessity of this operation.

\begin{figure}
    \centering
    \includegraphics[height=7.5cm, width=9cm]{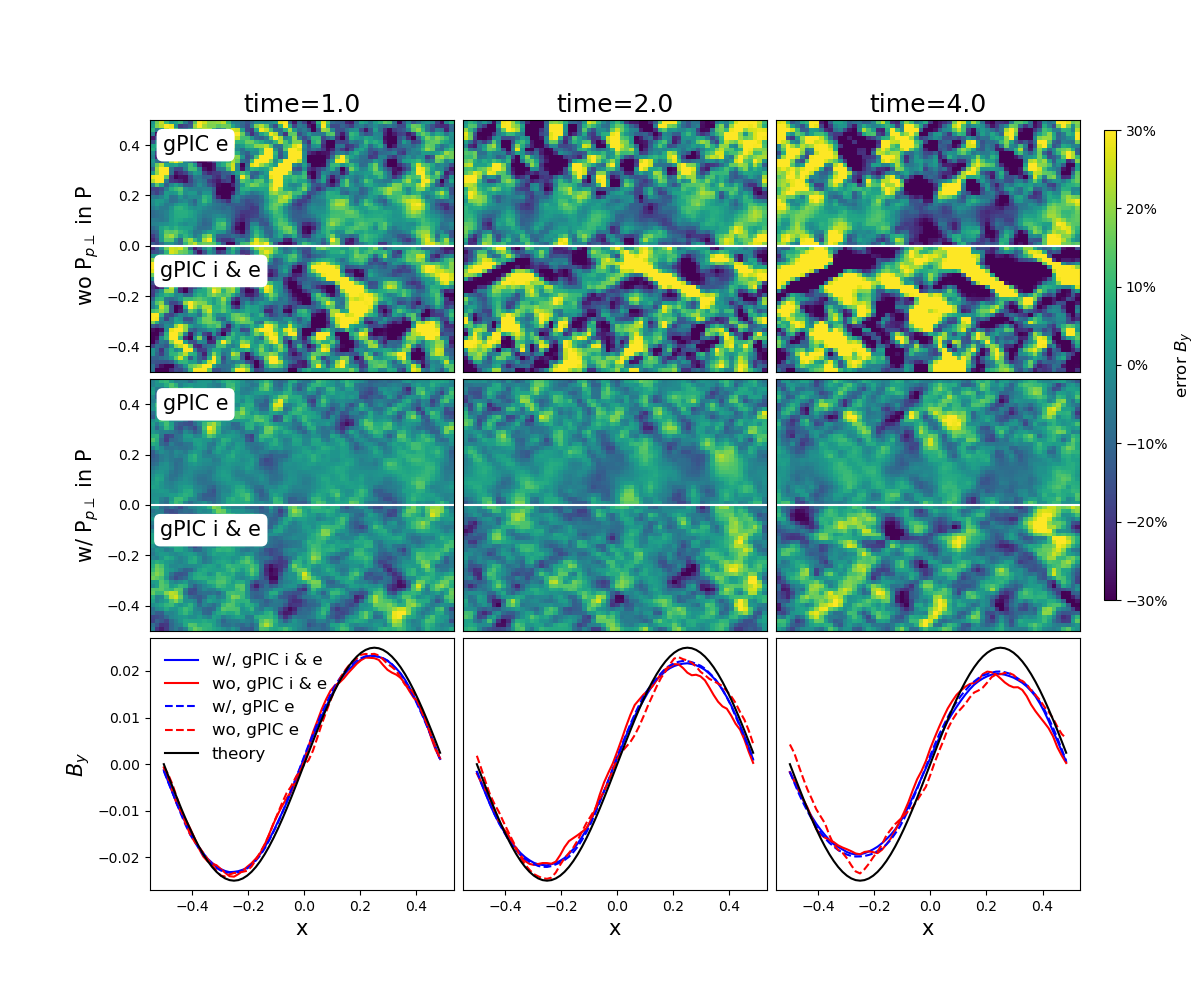}  
    \caption{{Simulation results for Alfvén wave propagation in horizontal direction with different initial particle distribution. First and Second row: initial particle distribution, titles describe the simulation size and whether include $\sfP_{p\perp}$ in $\sfP$; Third row: $B_y$ distribution at $t=5.0$}}
    \label{fig:noise}
\end{figure}

In the MHD-gPIC module, the gas in the MHD part is no longer charge neutral, thus the gas pressure here is from a combination of ions and electron fluids with generally unequal number/charge densities.
%From the initial development of this module, it was recognized that a significant portion of the plasma exhibits a non-thermal distribution. 
%However, the absence of prior knowledge regarding the non-thermal particle distribution can lead to an uneven spatial distribution. 
%Consequently, the pressure gradient may split into two components in an unpredictable manner, manifesting on both sides of the momentum-energy equations even when the total plasma distribution is smooth. 
%This splitting can introduce complexities in the interpretation of the equations and the simulation results.
{We do not make the assumption that $|n_{\rm fi}-n_{\rm fe}|\ll n_{\rm fi}$ to ensure broad range of applicability in our MHD-gPIC module. However, as the spatial distribution of particles can exhibit significant Poisson noise, this would make the thermal/non-thermal pressure spatially non-smooth, and directly evaluating the pressure gradient of individual components could significantly amplify the noise further. By incorporating $\sfP_{p\perp}$ into $\sfP$, we allow part of this noise amplification to be processed by the Riemann solver, which is expected to at least partially relieve the problem.}

In this test, we examine the propagation of Alfv\'en waves under an uneven particle distribution. {The particles share the same temperature as the thermal gas (for test purposes), and we keep the total number density for thermal and non-thermal plasmas constant, so that physically, the Alfv\'en waves should propagate as in a standard MHD gas.}
% We utilize the magnetic field configuration detailed in section \ref{sec:cpaw}. 
{The background magnetic field is aligned with the $x$-direction. 
A small perturbation with a wavelength equal to the $x$-extent of the simulation domain is added.}
The initial particle distribution is characterized by the following expressions
\begin{align}
    &n_\pe(x,y)=n_{\pe0} (1-\cos k_{\perp}y),\\
    &n_\text{pi}(x,y)=n_{\text{pi}0} (1+\cos k_{\perp}y),
\end{align}
where $n_{\pe0}$ and $n_{\text{pi}0}$ are predefined constants that determine the total number of electrons and ions, respectively. 
{The gPIC particles account for 40\% of the total plasma for each species (when included), with a local maximum of 80\% due to their non-uniform distribution.}
% This configuration models Alfvén wave propagation through a non-uniform background particle distribution, which varies perpendicularly to the magnetic field lines. 
The particle distribution is set to be uniform along the field lines.
%, as particles can move freely along these lines, resulting in an inherently even distribution.
The initial particle pressure is isotropic, and the gPIC component is initialized with the same temperature as the MHD component. 
To ensure a constant total pressure, the MHD gas pressure is adjusted according to the following equation
\begin{equation}
    \sfP_f(x,y)=\sfP_{f0}\xkh{1-\frac{n_{\pe}+n_\text{pi}}{2n_i}},
\end{equation}
where $n_i$ denotes the total ion density. 
This adjustment is crucial for maintaining a consistent total pressure, compensating for the fluctuations in particle densities.

The simulation is conducted in a 2D domain with a resolution of $(L_x/N_x, L_y/N_y)=(1/64,1/64)$.
The wave amplitude is set to $|B_\perp|=0.025B_0$.
We set $k_\perp=4\pi$ and initialized the gPIC electrons with $n_{\text{pe}0}=200$ and ions with $n_{\text{pe}0}=200$ when gPIC ions are also included.
Inevitably, the wave will experience damping due to noise arising from fluctuations in the particle distribution. 
To assess the simulation results, we employ the metric $\delta B_y/B_{y0}(t)=B_y/B_{y0}(t)-\sin (ky+\omega t)$ to evaluate the simulation result, where $B_{y0}(t)$ is derived from fitting $B_y$ data. 
In the first row of Figure \ref{fig:noise}, we present the evolution of $\delta B_y$ in a simulation that does not include {$\sfP_{p\perp}$ in MHD pressure}. 
The upper half of each panel depicts results from a simulation with only gPIC electrons, where regions of higher noise correspond to areas with a greater gPIC particle density. 
The lower half shows results from a simulation that includes both gPIC ions and electrons. 
By comparing these panels, it is evident that wave propagation in particle-dense regions is significantly affected by the presence of particles, and the wave can be disrupted by noise.
The second row of the figure illustrates the same simulation but with the inclusion of {$\sfP_{p\perp}$ in $\sfP$}. 
Here, despite the influence of noise, the Alfvén wave is able to propagate. 
In the third row, we plot the averaged $B_y$ along the $y$ direction for the aforementioned simulations, alongside the theoretical result. 
It is clear that the introduction of 
{$\sfP_{p\perp}$}mitigates noise, thereby facilitating the simulation of weaker wave propagation.

\begin{figure}
    \centering
    \includegraphics[height=10cm, width=7.5cm]{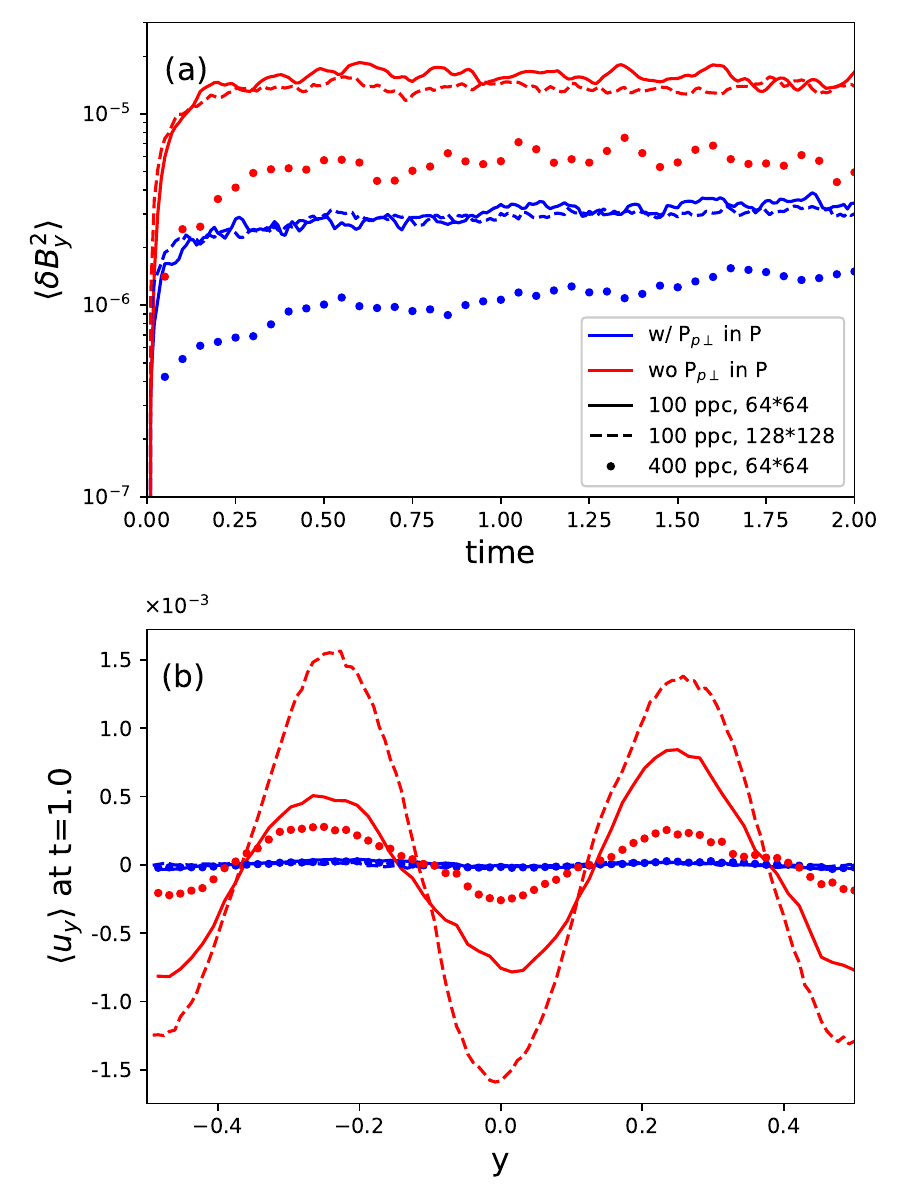}  
    \caption{(a) Time evolution of the magnetic fluctuation energy $\langle\delta B_y^2\rangle$ for different simulation configurations; 
    (b) Spatial distribution of $\langle u_y\rangle$ (the $x$-direction-averaged MHD gas velocity) along the $y$-direction at $t=1.0$.}
    \label{fig:truncation}
\end{figure}

By increasing the resolution or altering the average particle number per cell, we can quantify the noise level by calculating {$\braket{\delta B_y^2}$} as a function of time across various conditions. Selected simulation results featuring gPIC electrons are depicted in Figure \ref{fig:truncation}.
Upon comparing the simulation outcomes at different spatial resolutions in Figure \ref{fig:truncation} (a), we observe that the noise level remains unaffected by this parameter. 
However, the noise and damping effects can be further mitigated by increasing the number of particles per cell. 
Through a comparative analysis of simulations with different average particle numbers, we determined that more than 4 times the number of particles is required in simulations without the incorporation of $\sfP_{p\perp}$ to achieve a noise level comparable to that in simulations with it.

The underlying reason for this disparity is that the inclusion of $\sfP_{p\perp}$ effectively integrates an estimated particle pressure into the pressure calculations within the Riemann solver, thereby yielding more robust simulations.
In Figure \ref{fig:truncation} (b), we illustrate the distribution of the $y$-directional profile of $\braket{u_y}$ at $t=1.0$, where $\braket{u_y}$ represents the $x$-direction-averaged MHD gas velocity.
The data clearly indicates that the presence of {$\sfP_{p\perp}$ in $\sfP$} mitigates the truncation errors stemming from the non-uniform distribution of particles.
This characteristic facilitates the utilization of non-uniform initial particle distributions in simulations, allowing for a higher concentration of particles in regions of particular interest, which in turn enhances computational efficiency. 
Furthermore, it permits the introduction of additional particles into the simulation with a reduced impact on the stability of the code, providing greater flexibility in simulation design and execution.

\subsection{Magnetic Reconnection}\label{sec:reconnection}
Our final test considers magnetic reconnection, a fundamental plasma process crucial for generating high-energy particles, as a preliminary study toward more realistic astrophysical applications. The tests are conducted in 2D, and they enable us to reveal mechanisms for particle acceleration, test our static mesh refinement algorithm, and identify crucial areas for model enhancement in future work.

\begin{figure*}
    \centering
    \includegraphics[height=9cm, width=15cm]{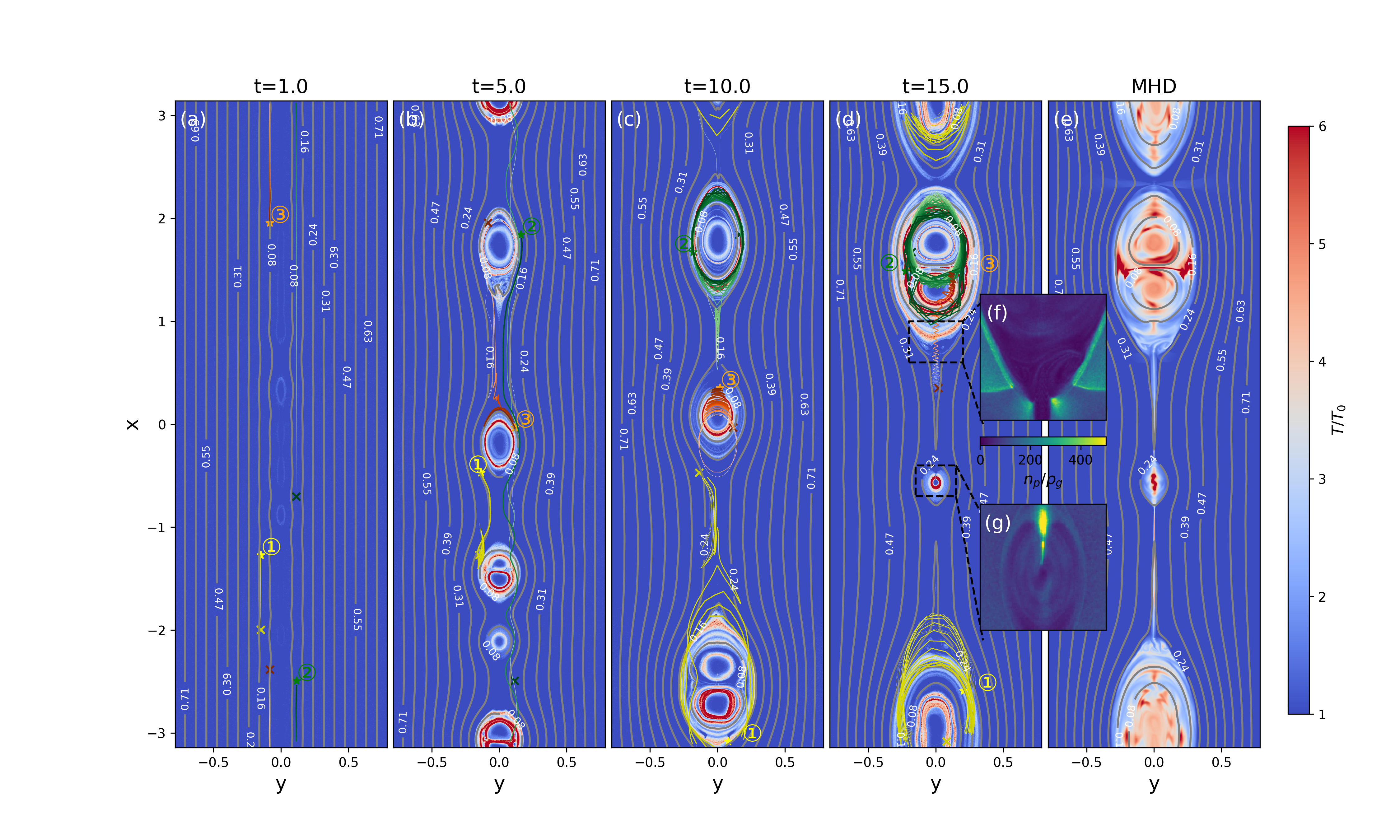}  
    \caption{Evolution of gas and {effective} particle temperature. (a–d) Sequential snapshots presenting the averaged particle energy, calculated as $(P_{\pe,\|}+2P_{\pe,\perp})/n_\pe$. The trajectories of three selected particles (labeled \ding{172}, \ding{173}, \ding{174}) are depicted; their initial positions ($t=0$ for panel (a)) are marked by "$\times$", and their positions at the respective snapshot times by  "$\star$". The connecting trajectories are rendered in different colors for each particle. (e) Depicts the gas temperature distribution at $t=15$, normalized to the initial temperature. In panels (a–e), the magnetic field configuration is depicted by contours of the vector potential $A_z$, with value labeled. (f, g) Illustrate the ratio of particle number density ($n_{\pe}$) to gas density $\rho_g$ within selected regions at $t=15$.}
    \label{fig:rec_field}
\end{figure*}

\begin{figure*}
    \centering
    \includegraphics[height=8cm, width=16cm]{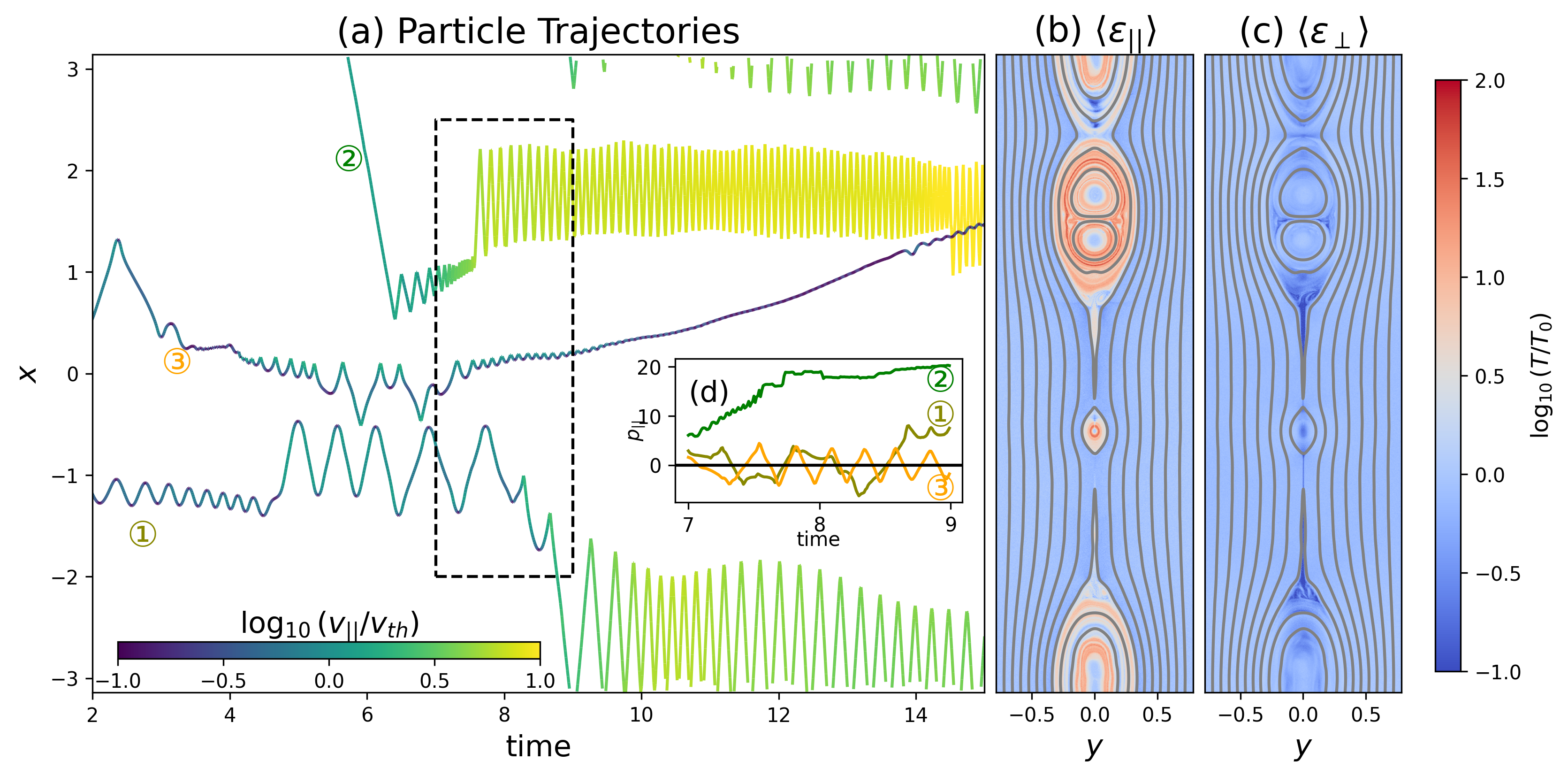} 
    \caption{Particle acceleration process from the uniform meshblock simulation (i). Panel (a) displays the $x$-coordinate evolution of selected particles, with color indicating their parallel velocity. Panel (b) and (c) show the averaged parallel ($\sfT_{\|\pe}/n_p$) and perpendicular ($\sfP_{\perp\pe}/n_p$) energy distributions of particles, respectively, calculated at $t=15$. Panel (d) illustrates the parallel momentum evolution for these specific particles during the time interval from $t=7$ to $t=9$.}
    \label{fig:rec_particle_energy}
\end{figure*}

\subsubsection{Simulation Setup}
The initial magnetic field configuration is given by
\begin{equation}
    \bm B(x,y)=B_0\tanh(y/\delta)\hat x+\sqrt{\frac{B_0^2}{\cosh^2 (y/\delta)}+B_g^2}\hat z,
\end{equation}
where $B_0=1$  is the reconnection field strength and $B_g=0.1$ is the guide field strength. 
The parameter $\delta$ representing the initial current sheet thickness, was set to approximately 5 grid sizes for the uniform mesh blocks. 
We consider a uniform initial plasma pressure $\sfP=0.125$ and density $\rho=1$.
To trigger the reconnection dynamics, we add an initial MHD velocity perturbation.
We initialize the particle distribution as a Maxwellian, consistent with the initial gas pressure and density.
Our model employs Ohmic resistivity with a coefficient $\eta_\Omega=10^{-5}V_AL_0$, which gives a system Lundquist number of $S=L_0V_A/\eta_\Omega=10^{5}$.

Following \cite{arnold2021}, the simulation domain is chosen to be $L_x\times L_y=2\pi L_0\times\pi/2L_0$, discretized with $N_x\times N_y=2048\times512$ grid cells on the root mesh.
We perform three simulations as follows:
\renewcommand{\theenumi}{(\roman{enumi})}
\begin{enumerate}
    \item \textbf{Electrons Only:} All particles represent electrons with 100 particles per cell. The initial electron thermal velocity is $3V_A$. No particle feedback or non-ideal electric field is included. No mesh refinement is applied.
    \item \textbf{Electrons with Refinement:} One level static mesh refinement is applied near the current sheet. The refined region contains 100 particles per cell, representing $10\%$ of the local electron population. 
    The electron thermal velocity matches case (i). 
    Particle feedback and non-ideal electric field are enabled.
    \item \textbf{Electrons + Ions with Refinement:} The MHD setup is identical to case (ii). The gPIC module includes both electrons and ions with 50 particles per species per cell, representing $5\%$ of the local electron and ion populations.
    The electron thermal velocity matches case (i), while the ion thermal velocity equals the MHD gas value $\sqrt{\sfP/2\rho}=0.25V_A$, yielding an ion-to-electron mass ratio of 144. 
    Particle feedback and non-ideal electric field are enabled.
\end{enumerate}

\subsubsection{Field Evolution and Particle Acceleration without Feedback}
Based on the simulation results from setup (i), we investigate particle acceleration during reconnection {by tracking representative particle trajectories}.

Initially, resistive damping of the magnetic field {in the current sheet leads to an imbalance of total pressure}, establishing a gas flow towards the current sheet. {As we do not apply Ohmic electric field to particles, there is minimum particle energy growth in this phase.} Together with random velocity perturbations, rapid magnetic reconnection is triggered, leading to the formation of a series of small plasmoids (Figure \ref{fig:rec_field}).
The left four panels of this figure show the averaged particle energy at different time
{snapshots}, while the rightmost panel displays the gas temperature distribution a $t=15$. As reconnection progresses, plasmoids continue to form and merge, converting magnetic energy into thermal energy — a process characteristic of fast plasmoid-mediated reconnection. {In the meantime, particles gain energy significantly.}

%{[I commented out the above. What you describe is inconsistent with what's seen in Fig. 7: gas temperature in the center of the plasmoids are not smaller, but generally higher than that in the periphery.]}

Particles are accelerated primarily through Fermi acceleration, a mechanism particularly effective during the contraction and merger of plasmoids. 
In our current model, the magnetic moment of each particle is assumed to be constant, and the resulting acceleration is primarily directed along the magnetic field lines, as the field is weakened during reconnection. {This can be seen in Figure \ref{fig:rec_particle_energy} (b) and (c), showing that for the overall particle population at late stages,}
the perpendicular energy distribution remains relatively unchanged, (primarily governed by the local magnetic field strength), while particle parallel energy experience significant gain.

The trajectories of two representative particles (\ding{172} and \ding{173}) are plotted in Figure \ref{fig:rec_field} (a)-(d), and the time
evolution of their $x$-coordinate, and parallel velocity (shown as color) and parallel momentum are shown in Figure \ref{fig:rec_particle_energy} (a) and (d).
Before particles significantly depart from the reconnection layer, they drift freely along the magnetic field lines towards the current sheet (Figure \ref{fig:rec_field} (a)). 
When particles approach the formed plasmoids, those with insufficient initial parallel momentum may become trapped between two adjacent plasmoids due to the mirror force. 
This trapping occurs because the magnetic field is compressed within the plasmoids and weakened in the regions between them (e.g., particle \ding{172} in Figure \ref{fig:rec_field} (b)). 
If these plasmoids subsequently move closer together, the trapped particles can experience significant acceleration during this process (e.g., particle \ding{172} between $t\sim2$ and $t\sim 4$ in Figure \ref{fig:rec_particle_energy} (a)). 
In contrast, particles with sufficiently high initial energy are less affected by the plasmoids and continue to move freely along the field lines (e.g., particle \ding{173} in Figure \ref{fig:rec_field} (b)). 
Upon entering the central reconnection region, particle acceleration becomes substantial, primarily through the Fermi mechanism, as illustrated by the parallel momentum evolution of particles \ding{172} and \ding{173} in Figure \ref{fig:rec_particle_energy} (d) \citep{Li2021}.

{Upon entering} a plasmoid, particles are free to traverse along magnetic field loop, forming a ring-like distribution of high-energy particles. {Their trajectories in Figure \ref{fig:rec_particle_energy} (a) then exhibit a zig-zag pattern. Further particle acceleration occurs primarily through plasmoid mergers, corresponding to a sudden relocation of the zig-zag pattern with increasing width in Figure \ref{fig:rec_particle_energy} (a). Such merger often incorporate other populations of particles into the larger plasmoids. Individual particle populations are then trapped in distinct magnetic flux loops, exhibiting as separate}
concentric rings of high-energy particles {(this is also because higher-order perpendicular drifts are negligible)}.
Particles located near the core of a plasmoid,
{have little chance to experience significant plasomid mergers, and hence}
do not experience significant acceleration. 
This results in a lower particle temperature at the plasmoid’s center compared to the surrounding regions. 

{We would also point out that the fact that particles are well confined in such concentric rings requires a particle integrator that is Galilean invariant. A direct adoption of the \citep{northrop1963} formula without corrections from our Equation (\ref{eq:dbdt}) applied in the particle momentum equation (\ref{eq:particle_momentum}) would mix up particles from different concentric rings, leading to additional artificial acceleration.}

{Figure \ref{fig:rec_spec} (a) shows the particle energy spectrum of all particles at different snapshots. Over time, a clear power law spectrum is developed with a power law index of $-2.8$.}
Due to the finite simulation domain size, the acceleration of the highest-energy particles is truncated once they enter the largest plasmoids.
{As particles trapped in plasmoids reside in}
surfaces of constant vector potential $A_z$.
We thus further partition the simulation domain based on $A_z$ and statistically analyze the particle energy spectra near different $A_z$ values at {$t=15$}. %{[Your figures tend to show t=15, but in the text you often refer to t=10.]}
The results are plotted in Figure \ref{fig:rec_spec} (b). 
Here, $A_z=0$ corresponds to the central plane at the simulation's outset. The magnetic field lines for different values of $A_z$ can be inferred from the contours labeled in Figure \ref{fig:rec_field} (e). 
{A comparison of particle energy distributions across different $A_z$ values shows that particles entering the plasmoids earlier do not reach a higher energy limit.
The most energetic particles are those on $A_z$ field lines that experience multiple plasmoid merging while avoiding the plasmoid core, thus enabling energy gain from post-merger field line contraction.}
%{[If this is your conclusion above, I don't see anything useful from this analysis, and suggest remove this part of discussion on the decomposition according to Az.]}

{From this test, we also identify subtle effects that could lead particle depletion or concentration around plasmoids, as shown in Figure \ref{fig:rec_field} (f), (g). The former case generally correspond to regions with very small $\bm B$, where particle parallel velocity is fastest. The latter case often correspond to the leading side}
of a plasmoid along the direction of motion ($\bm u\cdot\bm\kappa<0$). We find that the $\nabla B$ term in particle motion can cause significant particle acceleration and deceleration during their orbit in the loop, exemplified by the trajectory of particle \ding{174} shown in Figures \ref{fig:rec_field} and \ref{fig:rec_particle_energy}. Numerical truncation errors inherent in the particle pusher can cause particles to {lose energy on average and} become trapped or settle in these locations.
%{[Originally, you first quote Arnold+21, saying our results are different from theirs. Then you say all the stuff above. This sounds like they did everything right, while we have many problems. Think about it!]}

\subsubsection{Mesh Refinement and Particle Feedback}\label{sec:mesh_refinement}

\begin{figure}
    \centering
    \includegraphics[height=8cm, width=8cm]{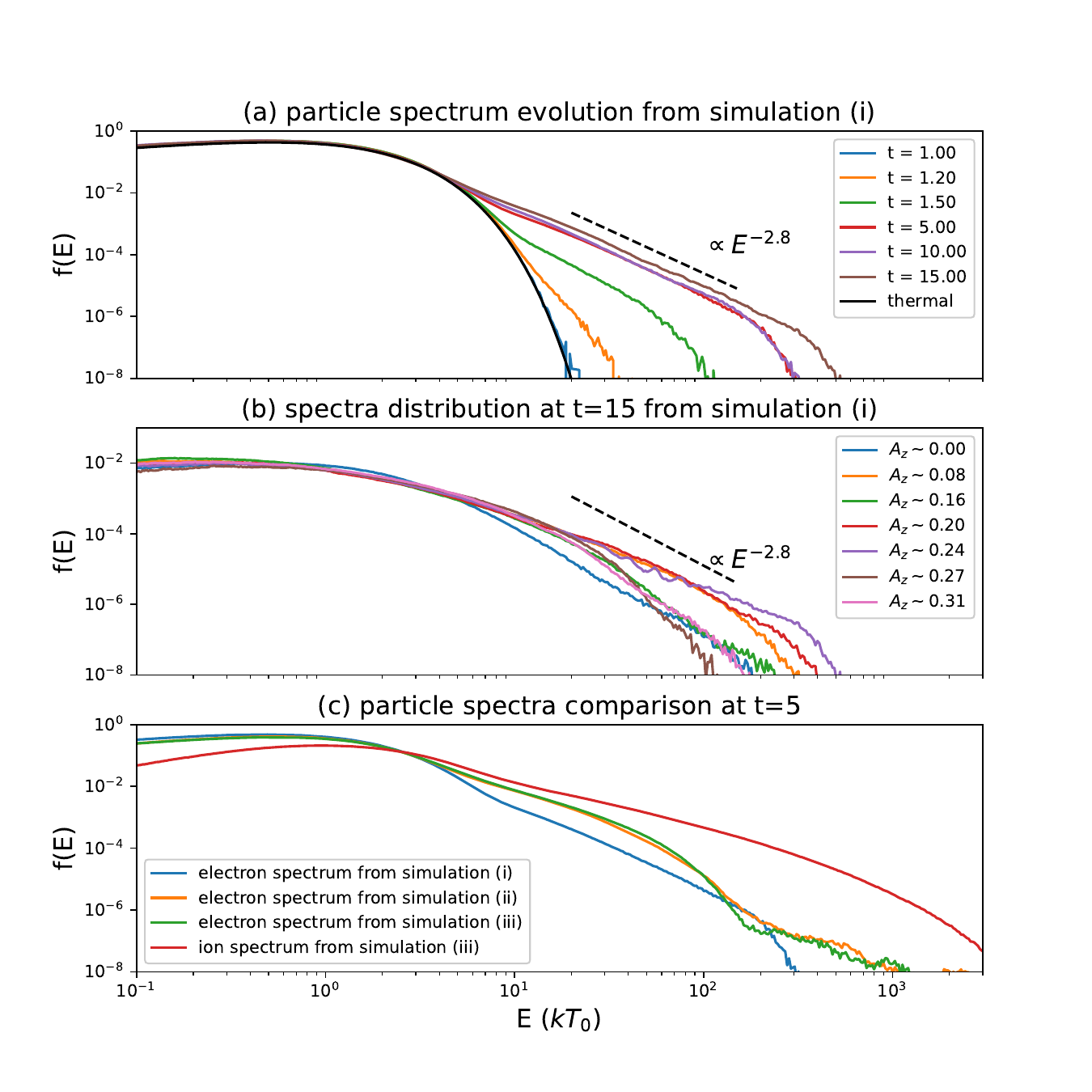} 
    \caption{
        (a) Time evolution of the particle energy spectrum during magnetic reconnection. The black line denotes the initial Maxwellian distribution. The spectrum at the final time develops a power-law range with a spectral index of $-2.8$.
        (b) Energy spectrum of electrons associated with different magnetic field lines at $t=15$.
        (c) Comparison of particle energy spectra from different simulation configurations at $t=5$.
    }
    \label{fig:rec_spec}
\end{figure}

\begin{figure*}
    \centering
    \includegraphics[height=12cm, width=16cm]{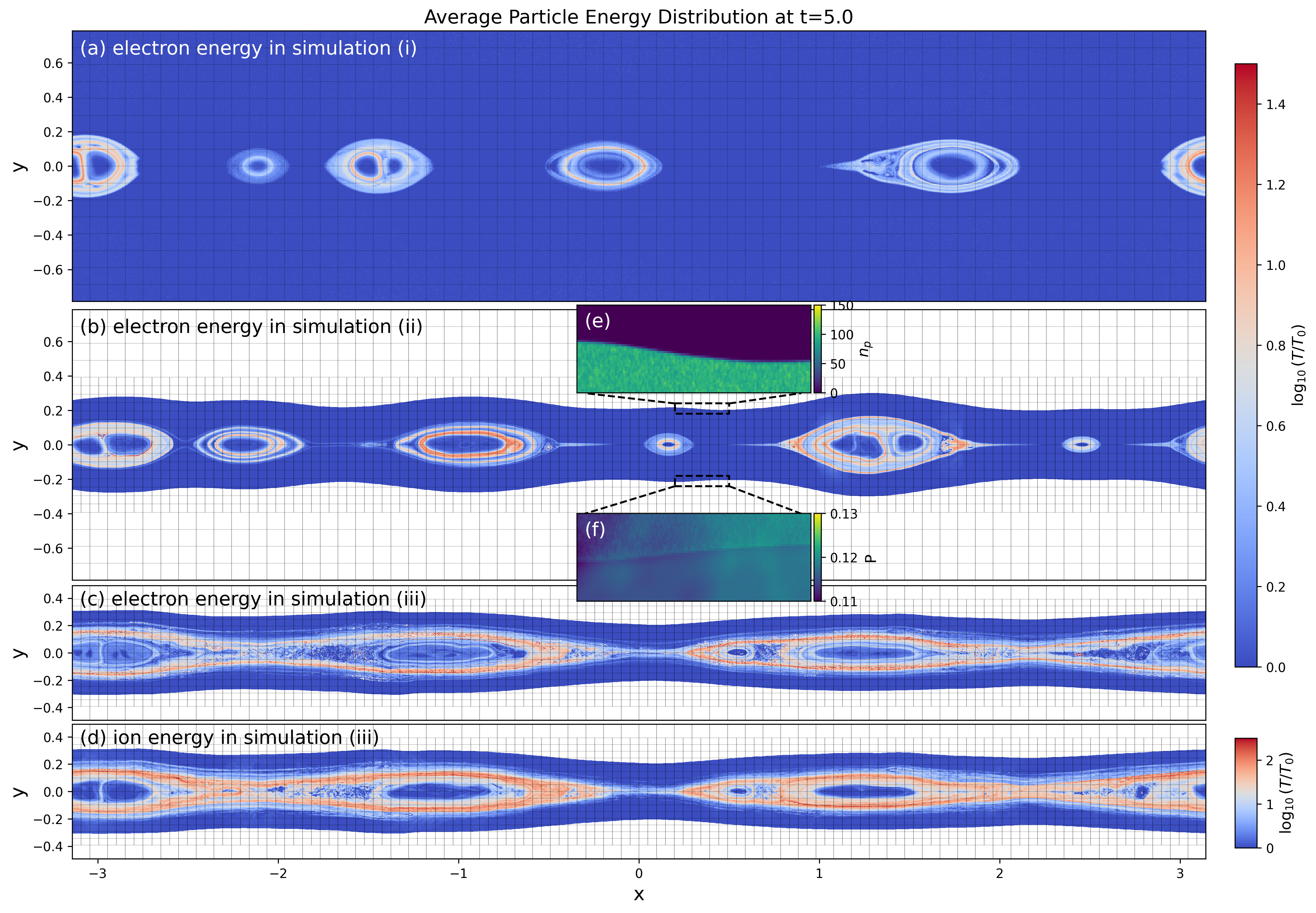} 
    \caption{Energy distributions of particles at time $t=5$ for different simulation setups. The black lines represent the mesh block boundaries, with each block containing $32\times32$ grids. show the electron energy distribution for the gPIC component in simulations (i), (ii), and (iii), respectively. Panel (d) displays the ion energy distribution for the gPIC component from simulation (iii). Panel (e) presents the particle number density within the selected region. Panel (f) shows the pressure $\sfP$ in the selected region.}
    \label{fig:rec_field_mr}
\end{figure*}

Figure \ref{fig:rec_field_mr} (a) and (b) illustrate the mesh structure for simulations (i) and (ii), where the black lines represent the boundaries of the mesh blocks, each comprising $32\times32$ grids. Mesh refinement in (ii) enables the study of finer structures. However, if the total particle number remains constant, the particle number per cell decreases as the grid size is reduced, potentially increasing noise due to particle fluctuations. 
To mitigate particle noise while balance computational efficiency, we limit particle initialization to the central region defined by $-0.25<y<0.25$, with particle number density to 100 particles per cell as particles far from the reconnection region are not significantly accelerated during the simulation.

{By comparing Figure \ref{fig:rec_field_mr} (a) and (b), we find that} in the absence of gPIC ions, the electron acceleration dynamics and plasmoid morphology in {simulation (ii) with both mesh refinement and particle feedback} align well with those {in simulation (i) (no refinement, no feedback)}.
{It is worth noting that at the boundary of the region with initially injected particles, there is a sharp discontinuity in particle number density, but the distribution in $\sf P$ remains smooth (Figure \ref{fig:rec_field_mr} (e) and (f)). This reinforces our choice in the MHD-gPIC formulation to mitigate particle noise.}
Despite the similarities in large-scale features, a main discrepancy emerges in the electron energy spectrum. 
Specifically, the mesh refinement simulation produces a spectrum that extends to considerably higher energies than observed in simulation (i), but it fails to exhibit a coherent power-law distribution (see Figure \ref{fig:rec_spec} (c)). 

Turning to simulation (iii) which further incorporate ions, we find that the significantly greater mass of ions leads to more pronounced acceleration during reconnection, as shown in
Figure \ref{fig:rec_spec} (c). The electron spectrum remains consistent with simulation (ii), whereas a higher fraction of ions are strongly accelerated with a harder spectrum, with the ion spectra exhibit much higher energy cutoffs. 
Despite representing a small fraction (5\%) of the initial particle population, the gPIC ions exert a considerable influence on the MHD component due to their efficient acceleration and relatively low velocities, leading to significant feedback. 
This effect is visually apparent in the increased turbulence across the current sheet (Figure \ref{fig:rec_field_mr} (c), (d)), the emergence of granular structures indicative of localized ion density, and the more elaborate, extended morphology of plasmoids. 

\subsubsection{Comparison with {\it kglobal} Results}\label{ssec:rec_compare}

Our test results can be compared with those from the {\it kglobal} reconnection simulations by \citet{arnold2021} on electron acceleration and \citet{Yin_Ion} who further included ions. The results are consistent at qualitative level, including plasmoid formation, Fermi acceleration, and more significant ion acceleration compared to electrons, but quantitative differences emerge in spectral properties.

Relative to \textit{kglobal}, our simulations yield lower maximum particle energies and a different spectral index. 
We attribute these differences, at least in part, to numerical origins, as discussed in section \ref{sec:comparison}. 
%The most likely explanation is 
We speculate that particles in \textit{kglobal} are not tightly confined within plasmoids and can therefore undergo further acceleration. 
This arises likely because {\it kglobal} model %adopts the classic \citet{northrop1963} formulation
does not well preserve Galilean invariance and includes particle spatial diffusion, which allow particles to drift away from confinement regions.
These numerical effects inadvertently promote particle acceleration by reducing particle trapping.

Such technical differences, besides the difference in the base MHD framework between the two models, preclude direct and fair comparison between our results with theirs. We therefore stress that the outcome of particle acceleration in magnetic reconnection simulations can be subject to numerical subtleties and should be treated with caution.

\section{Summary and Discussion}\label{sec:summary}
This paper presents the MHD-gPIC module implemented in the \verb|Athena++| code, combining magnetohydrodynamics (MHD) and particle-in-cell (PIC) methods for multi-scale plasma kinetic simulations. 
The module treats a fraction of the plasma (hot electrons in this study) as particles using the guiding center approximation, while the rest of the plasma is modeled with MHD equations, with carefully derived source terms from particle backreaction. Compared to the original MHD-PIC formulation \citep{bai2015}, the guiding center approximation substantially alleviates the issue of scale separation in applications where gyro-resonance is known to be sub-dominant. It is designed to enable multi-scale simulations by capturing particle effects at small scales while maintaining the large-scale features of MHD.
%issue but may require consideration of relativistic effects for electrons in some cases.

The main innovations in our formulation and algorithm include:
%\begin{itemize}
    %\item 
(1). The particle momentum equation incorporates a missing term in conventional guiding-center equations to ensure the integrator is Galilean invariant. This is critical for confining particles to moving plasmoids during magnetic reconnection.
    %\item 
(2). Our particle backreaction source terms are derived so that they permit a non-negligible fraction non-thermal particles, significantly broadening the model's applicability.
    %\item 
(3). We augment fluid pressure to further include the perpendicular pressure of gPIC particles ${\sf P}_{p\perp}$. This approach leverages the Riemann solver's dissipation which significantly reduces Poisson noise from particle pressure gradients.
%\end{itemize}

The module, built upon the \verb|Athena++| MHD-PIC infrastructure \citep{Sun2023}, is compatible with static and adaptive mesh refinement (SMR/AMR) and fully supports load balancing in parallel computations. The overall time evolution is second-order accurate. Despite interpolating more variables per particle for interpolation and deposits, the computational overhead remains modest (approximately 10--18\% per particle relative to MHD-PIC module).
%This cost is offset by the ability to take significantly larger numerical time steps under the guiding-center approximation.
%\xb{[Add the highlights of the formulation and algorithm: works for non-negligible np/n, Galilean invariant, add Pperp to P for noise mitigation, compatible with SMR/AMR and load balancing.]}
%The algorithm achieves second-order accuracy in time and conserves total momentum and energy.
%The MHD-gPIC module inherits the vectorization and parallel perfomance from the MHD-PIC module \citep{sun2023}.
%(Figure \ref{fig:benchmark}).

We validate the accuracy and performance of the MHD-gPIC module with a series of benchmark tests.
%The trajectory test with a 2D magnetic field loop demonstrates that the guiding center approximation accurately captures the particle trajectory when compared to the MHD-PIC module (Section \ref{sec:loop}).
%The circularly polarized Alfvén wave test verifies the correct implementation of the particle source terms in the MHD momentum equation (Section \ref{sec:cpaw}).
%The electric acoustic wave test validates the electric field calculation in the MHD-gPIC method (Section \ref{sec:eaw}).
%The SMR simulation in magnetic reconnectin test converge to the same result in a uniform grid (Section \ref{sec:mesh_refinement}), validating the interpolation/deposit scheme across refinement boundaries.
These tests demonstrate that this module can accurately integrate particle trajectory and capture its momentum, energy and electromagnetic feedback with reduced Poisson noise. As a particular example, by initializing particles with a distribution function that mimics background thermal plasma, the coupled particle-gas system behaves exactly as pure gas with added particle density.

As a primary target application, we conducted three magnetic reconnection tests with increasing physical ingredients to study particle acceleration. 
%Our simulations, incorporating mesh refinement and particle feedback, 
These simulations successfully demonstrated that particles are primarily accelerated via the Fermi mechanism during plasmoid formation and mergers. Significant energy gain was observed, particularly in the parallel direction, leading to a pronounced pressure anisotropy (Figure~\ref{fig:rec_particle_energy}). The simulation without particle feedback form an extended power-law energy spectrum. Incorporating particle feedback, particles can be accelerated to even higher energies, but the energy spectra no longer exhibit a coherent power-law
%spectrum was not formed under our current model setup with a fixed particle fraction and idealized magnetic moment conservation 
(Figure~\ref{fig:rec_spec}). We also find that the inclusion of gPIC ions, despite their small initial number fraction, significantly influences the MHD environment through feedback, leading to enhanced turbulence and more elaborate plasmoid structures. The ions are also more significantly accelerated than electrons.
%These tests confirm the robustness of our implementation, showing that the morphology and large-scale acceleration dynamics were consistent across simulations with and without mesh refinement/feedback.

Our test simulations of magnetic reconnection confirm the robustness of our implementation of the MHD-gPIC altorighm, while the idealized setup point to areas that call for future improvements, primarily in the following two aspects.
%The absence of a fully developed power-law spectrum in our reconnection test with particle feedback, when compared to certain other models (e.g., \citealt{arnold2021}), underscores interesting physical and numerical differences that merit further exploration within our MHD-gPIC framework. We identify two plausible factors that may contribute to these differences and propose corresponding directions for future investigation:
%\begin{itemize}
    %\item Particle injection. 
First, our test simulations treat particles as pre-existing with a uniform Maxwellian distribution. In reality, non-thermal particles are expected to emerge only at
    %In realistic reconnection scenarios, particles are primarily injected upon entering the diffusion region around X-points and during plasmoid mergers. 
    %A more physical, \textit{on-the-fly} injection prescription, triggered by identifying these 
    specific locations and events (e.g., \citealt{French2025} for relativistic reconnection) through kinetic processes, and our formalism are designed to capture subsequent processes of acceleration. More realistic study should properly identify potential injection locations
    (e.g., \citealt{Wang2023}), with a prescription for initially injected particle spectrum.
    %could drastically improve the realism of the resulting particle energy spectrum and potentially enhance the simulated reconnection rate. This would ensure that the computational resources focus on particles that are most likely to be accelerated.
    %\item Physical Diffusion. 
Second, the guiding-center approximation, which assumes perfect magnetic moment conservation, neglects gyro-scale wave-particle interactions which leads to pitch angle diffusion. Moreover, the coarse resolution also neglects potential particle spatial diffusion due to field line wandering at sub-grid scale. Both effects depend on sub-grid turbulence not known {\it a priori}, but may be modeled by prescriptions \citep{Jokipii1971, Kong2019, Chen2024} which potentially lead to more efficient particle acceleration.
Incorporating such improvements in a physics-driven manner would be crucial
%aimed at making injection and diffusion more physics-driven rather than numerically prescribed, are expected to 
bridge the connection between kinetic and macroscopic scales toward a true multiscale characterization of particle acceleration process during non-relativistic magnetic reconnection.

\section*{Acknowledgements}

We thank Marc Swisdak and Zhiyu Yin for constructive discussions on the {\it kglobal} model, and Fan Guo, Shinsuke Takasao and Yulei Wang for helpful exchanges. This work is supported by the National Science Foundation of China under grant No. 12325304, 12342501.

\appendix
\section{Guiding Center Approximation}\label{appendix:GC_equations}
We use $\C=1$ in this section.
The basic figure is a particle with mass $m$ and charge $q$ of the MHD gas moves in electro-magnetic field $F^{\mu\nu}$.
The momentum equation of this particles is
\begin{equation}
    \dd{v^\mu}\tau=\frac qm F^{\mu\nu} v_\nu,
\end{equation}
where $v^\mu\equiv \D x^\mu/\D\tau$ is the 4-velocity.

The electro-magnetic field is nearly ideal, which means $F^{\mu\nu}$ can be decomposed as
\begin{equation}
    F^{\mu\nu}=\epsilon^{\mu\nu\alpha\beta}u_\alpha b_\beta-E_\|\hat b^\mu u^\nu+E_\|\hat b^\nu u^\mu,\label{eq:EMfield}
\end{equation}
where $u^\mu$ a 4-velocity whose spatial components correspond to the perpendicular-$b$ part of the MHD gas velocity, i.e., $u^\mu u_\mu=-1$ and $b^\mu u_\mu=0$, $b^\mu$ is the magnetic field 4-vector \citep{Gammie2003}, $\hat b^\mu\equiv b^\mu/\sqrt{b^\nu b_\nu}$ is the unit vector of magnetic direction, and $E_\|$ is the non-ideal parallel electric field which satisfies $E_\|\ll b$.
We use $b\equiv \sqrt{b^\nu b_\nu}$ to represent magnetic field strength hereafter.
The non-ideal perpendicular is negligible compare with the ideal electric field.
In this situation, the pure imaginary eigenvalues of $qF^{\mu\nu}/m$ are $\pm i\omega$ where $\omega\equiv qb/\gamma m$. 
And we use $\delta^\mu,\ \sigma^\mu$ to represent eigenvectors correspond to $\pm i\omega$ \citep{Trent2024}, they satisfy 
\begin{align}
    &\delta_\mu=\sigma_\mu^*,\\
    &\delta_\mu\sigma^\mu=1,\\
    &\delta_\mu\delta^\mu=\sigma_\mu\sigma^\mu=0,\\
    &\delta_\mu u^\mu=\delta_\mu b^\mu=\sigma_\mu u^\mu=\sigma_\mu b^\mu=0,\\
    &\delta_\mu\sigma_\nu+\delta_\mu\sigma_\nu=g_{\mu\nu}+u_\mu u_\nu-\hat b_\mu\hat b_\nu,\label{eq:delta_sigma}
\end{align}
the last equation can be found directly in MHD gas co-moving coordinate.
Therefore, $(u^\mu,\hat b^\mu,\delta^\mu,\sigma^\mu)$ make up a set of basis of the Minkowski spacetime.

In this basis, the particle velocity can be decomposed as
\begin{equation}
    v^\mu=\gamma_0 u^\mu+v_\|\hat b^\mu+v_\perp e^{i\omega\tau+i\phi} \delta^\mu+v_\perp e^{-i\omega\tau-i\phi} \sigma^\mu,\label{eq:v_particle}
\end{equation}
where $\gamma_0$ is the Lorentz factor in MHD gas co-moving coordinate, $v_\|$ and $v_\perp$ are corresponding parallel and perpendicular momentum, $\phi$ is some constant phase.
Obviously, $\gamma_0^2=1+v_\|^2+2v_\perp^2$.
We define the guiding center position $X^\mu$ from real position $x^\mu$ by
\begin{equation}
    X^\mu\equiv x^\mu+\frac{mF^{\mu\nu}v_\nu}{qb^2},\label{eq:GC_X}
\end{equation}
which corresponds to equation (\ref{eq:x_gk}).
The guiding center velocity is
\begin{equation}
\begin{aligned}
V^\mu&\equiv\dd{X^\mu}\tau=v^\mu+\frac{F^{\mu\nu}F_{\nu\lambda}}{b^2}v^\lambda+\frac {m}{q}\dd{F^{\mu\nu}/b^2}\tau v_\nu=\underbrace{\gamma_0 u^\mu+v_\|\hat b^\mu}_{V_{(0)}^\mu}+\underbrace{\frac mq v_\nu v_\lambda\nabla^\lambda\frac{F^{\mu\nu}}{b^2}}_{V_{(1)}^\mu} ,
\label{eq:GCvelocity}
\end{aligned}
\end{equation}
where we used equation (\ref{eq:EMfield}) and $\D/\D\tau\equiv v_\mu\nabla^\mu$ in above derivation and omitted the $\mathscr{O}(E_\|^2/b^2)$ terms.
% The first two terms make up the 0th-order GC velocity 
% \begin{equation}
% V_\text{(0)}^\mu=\gamma_0 u^\mu+v_\|\hat b^\mu.
% \end{equation}
By dropping non-ideal electric field in $F^{\mu\nu}$,  averaging the high frequency term in $V^\mu_{(1)}$ with
\begin{equation}
\begin{aligned}
    \braket{v_\nu v_\lambda}=&\xkh{\gamma_0^2+v_\perp^2}u_\nu u_\lambda+\gamma_0 v_\|(u_\nu\hat b_\lambda+\hat b_\nu u_\lambda)+\xkh{v_\|^2-v_\perp^2}\hat b_\nu\hat b_\lambda+v_\perp^2g_{\nu\lambda},\\
\end{aligned}\label{eq:gyroaverage}
\end{equation}
and using
\begin{align*}
    &u_\nu \nabla^\lambda\frac{F^{\mu\nu}}{b^2}=-\frac{F^{\mu\nu}}{b^2}\nabla^\lambda u^\nu, \\
    &\hat b_\nu \nabla^\lambda\frac{F^{\mu\nu}}{b^2}=-\frac{F^{\mu\nu}}{b^2}\nabla^\lambda \hat b^\nu,\\
    &g_{\nu\lambda}\nabla^\lambda\frac{F^{\mu\nu}}{b^2}=-2F^{\mu\nu}\frac{\nabla_\nu b}{b^3}+\frac{J^\mu}{b^2} =-F^{\mu\nu}\frac{\nabla_\nu b}{b^3}+\frac1{b^2}\xkh{J_\|\hat b^\mu-F^{\mu\beta}\kappa_\beta+F^{\mu\nu}\DD{ u_\nu}t},
\end{align*}
where we used Maxwell's equation $\nabla_\nu F^{\mu\nu}=J^\mu$, and $\kappa_\beta\equiv\hat b^\nu\nabla_\nu\hat b^\beta$ is the magnetic field curvature,
the 1st-order GC velocity includes
\begin{equation}
    V_\text{(1)}^\mu=\underbrace{V^\mu_\text{Speiser}}_{\frac{\mu_0 J_\|\hat b^\mu}{qb}}+\underbrace{V^\mu_{\kappa}}_{\frac{mv_\|}{q b^2}F^{\nu\mu}\xkh{v_\|\kappa_\nu+\gamma_0\DD{\hat b_\nu}t}}+\underbrace{V^\mu_{\nabla b}}_{\frac{\mu_0 }{qb^2}F^{\nu\mu}\nabla_\nu b}+\underbrace{V^\mu_\text{else}}_{\frac{F^{\nu\mu}}{\omega b}\gamma_0\dd{u_\nu}t}, \label{eq:GC_V1}
\end{equation}
where we defined the magnetic moment $\mu_0\equiv mv_\perp^2/b$,  $\text{D}/\text{D}t\equiv u^\mu\nabla_\mu$ and $\nabla_\|=\hat b^\mu\nabla_\mu$.
The first term is the Speiser motion velocity \citep{Speiser1965}, the second term is the curvature drift velocity and the third term is the gradient-$B$ drift velocity.

The GC momentum equation can be found from equation (\ref{eq:GCvelocity})
\begin{equation}
    \begin{aligned}
        \dd{V^\mu}\tau&=\dd{}\tau\xkh{v^\mu+\frac{F^{\mu\nu}F_{\nu\lambda}}{b^2}v^\lambda+\frac mq\dd{F^{\mu\nu}/b^2}\tau v_\nu}=\frac qmF^{\mu\nu}V_\nu+\dd{F^{\mu\nu}}\tau\frac{F_{\nu\lambda} v^\lambda}{b^2}+\xcancel{\dd{V^\mu_{(1)}}\tau},
    \end{aligned}
\end{equation}
here we omit the last term of equation to get the $\omega^0$ order terms, and gyro-averaged result of the second term is
\begin{align}
    \braket{\dd{F^{\mu\nu}}\tau\frac{F_{\nu\lambda}v^\lambda}{b^2}}&=\frac{\braket{v^\sigma v^\lambda}}{b^2}F_{\nu\lambda}\nabla_\sigma F^{\mu\nu}=\frac{v_\perp^2}{b^2}F_{\nu\sigma}\nabla^\sigma F^{\mu\nu},
\end{align}
where we used equation (\ref{eq:gyroaverage}).
Using Maxwell's equation
\begin{equation*}
    \nabla^\sigma F^{\mu\nu}+\nabla^\mu F^{\nu\sigma}+\nabla^\nu F^{\sigma\mu}=0,
\end{equation*}
and
\begin{equation*}
    F_{\nu\sigma}\nabla^\sigma F^{\mu\nu}=F_{\nu\sigma}\nabla^\nu F^{\sigma\mu},
\end{equation*}
we can find
\begin{equation}
    \braket{\dd{F^{\mu\nu}}\tau\frac{F_{\nu\lambda}v^\lambda}{b^2}}=-\frac{v_\perp^2}{2b^2}F_{\nu\sigma}\nabla^\mu F^{\nu\sigma}=-\frac{v_\perp^2}{b}\nabla^\mu b.
\end{equation}
The final momentum equation is
\begin{equation}
    \dd{V^\mu}\tau=\frac qmF^{\mu\nu} V_\nu-\frac{\mu_0}{m}\nabla^\mu b.\label{eq:GC_ME}
\end{equation}
The mass energy equation for guiding center motion is
\begin{equation}
    V^\mu V_{\mu}+\frac{2\mu_0 b}m=-1,
\end{equation}
we can find $\mu_0$ is a constant by combining the above two equations.

Based on equation (\ref{eq:GC_V1}) and (\ref{eq:GC_ME}), we can find the parallel momentem equation of guiding center as
\begin{equation}
    \dd{V_\|}\tau=\dd{V^\mu \hat b_\mu}\tau=V^\mu\dd{\hat b_\mu}\tau+\frac qm E_\|-\frac{\mu_0}m\nabla_\|b.
\end{equation}
Similarly, the energy equation is
\begin{equation}
    \dd{V^0}\tau=\frac{q}{m}E_\| v_\|+\frac{\mu_0}{m}\DD bt+v_\|u_\mu\xkh{v_\|\kappa^\mu+\gamma_0\DD{\hat b^\mu}t}+\gamma_0^2\dd{u^0}t,\label{eq:GC_E}
\end{equation}
where we assume $u^0\approx1$ in above equation.
The momentum energy equations are used in Section \ref{sec:particle}. 
\section{GRMHD-gPIC model}\label{appendix:GRMHD-gPIC}
Similar to Section \ref{sec:fluid_closure}, fluid variables can also link to individual particles in GRMHD.
For each particle $k$, its worldline $x_k^\mu$ is a function of its proper time $\tau_k$.
The fluid density $\rho(x^\nu)$ can be expressed by
\begin{equation}
    \rho(x^\nu) \equiv\sum_k\int_{-\infty}^\infty\frac{\delta^4(x^\nu-x^\nu_k(\tau_k))}{\sqrt{|g|}}\D\tau_k,
\end{equation}
where $|g|$ is the determinant of the local metric $g^{\mu\nu}$.
Similarly, the fluid 4-velocity is
\begin{equation}
    \rho u^\mu =p^\mu(x^\nu)\equiv\sum_k\int_{-\infty}^\infty\frac{\delta^4(x^\nu-x_k^\nu(\tau_k))}{\sqrt{|g|}}\dd{x_k^\mu}{\tau_k}\D\tau_k,
\end{equation}
and the stress tensor is
\begin{equation}
    \sfT^{\mu\nu}(x^\lambda)\equiv\sum_k\int_{-\infty}^\infty\frac{\delta^4(x^\lambda-x_k^\lambda(\tau_k))}{\sqrt{|g|}}\dd{x_k^\mu}{\tau_k}\dd{x_k^\nu}{\tau_k}\D\tau_k.
\end{equation}
It is easy to proof $p^\mu_{\ \ ;\mu}=0$ and 
\begin{equation}
    \sfT^{\mu\nu}_{\quad;\mu}=\frac qm F^\nu_\mu p^\mu= \sum_k\int_{-\infty}^\infty\frac{\delta^4(x-x_k(\tau_k))}{\sqrt{|g|}}\dd{^2x_k^\nu}{\tau_k^2}\D\tau_k.
\end{equation}
If we apply guiding center approximation to part of particles, the momentum-energy closure equation of the gPIC part is
\begin{equation}
    {\sfT}^{\mu\nu}_{p\ ;\mu}=\sum_{k \in \gPIC}\int_{-\infty}^\infty\frac{\delta^4(x-x_k(\tau_k))}{\sqrt{|g|}}\dd{^2x_k^\nu}{\tau_i^2}\D\tau_k,
\end{equation}
from equations (\ref{eq:v_particle}) and (\ref{eq:GC_X})
\begin{equation}
\begin{aligned}
x_k(\tau_k)&=X_k(\tau_k)+\delta x_k(\tau_k)\approx X_k(\tau_k)+i\rho_ie^{i\omega_k\tau_k+i\phi_k}\sigma-i\rho_ke^{-i\omega_k\tau_k-i\phi_k}\delta,
\end{aligned}
\end{equation}
where the Larmor radius is $\rho_k\equiv q_kv_{i\perp}/m\omega_k$.
Using 
\begin{equation}
    \delta^4(x-x_k(\tau_k))=\delta^4(x-X_k(\tau_k))-\delta x_k^\mu\nabla_\mu\delta^4(x-X_k(\tau_k)),
\end{equation}
the result of energy-momentum equation of the gPIC part is
\begin{equation}
\begin{aligned}
    \sfT^{\mu\nu}_{p\ ;\mu}&=\sum_k\int_{-\infty}^\infty\frac{\delta^4(x-X_k(\tau_k))}{\sqrt{|g|}}\dd{^2X_k^\nu}{\tau_k^2}\D\tau_k+\nabla_\mu\sum_k\int_{-\infty}^\infty\frac{\delta^4(x-X_k(\tau_k))}{\sqrt{|g|}}\rho_k^2\omega_k^2(\sigma^\mu\delta^\nu+\delta^\mu\sigma^\nu)\D\tau_k.
\end{aligned}\label{eq:GR_closure}
\end{equation}
In near-ideal MHD, with equation (\ref{eq:delta_sigma}), equation (\ref{eq:GR_closure}) degenerates to equation (\ref{eq:momentum_closure}) and (\ref{eq:energy_closure}).

Following the deduction in Section \ref{sec:gas_dynamics}, the total energy-momentum equation is
\begin{equation}
    \sfT^{\mu\nu}_{\MHD;\mu}=\nabla^\nu\sfT_{p\perp}-\sfT_{p\ ;\mu}^{\mu\nu},
\end{equation}
where $\sfT_\MHD=\sfT_{f}+{\sfT}_{p\perp}+\sfT_\text{EM}$ and $\sfT_{p\perp}\equiv \sfT^{\mu\nu}_p(\delta_\mu\sigma_\nu+\sigma_\mu\delta_\nu)/2$ is the perpendicular pressure of gPIC part.
The above equation together with Maxwell's equations \citep{Gammie2003, White2016} and GC equation (\ref{eq:GC_ME}) forms the GRMHD-gPIC model.

\section{Derivation of Selected Equations}\label{appendix:deduction}
\subsection{Equation (\ref{eq:dbdt_MHD})}
To derive the expression for the time derivative of $\bm B$, we begin with the fundamental equation governing the magnetic field’s evolution in ideal MHD
\begin{equation}
    \p_t\bm B=\nabla\times\xkh{\bm u\times\bm B}=-\bm u\cdot\nabla\bm B+ B\nabla_\|\bm u-\bm B\nabla\cdot\bm u,
\end{equation}
where the electric field $\bm E$ is approximated by $\bm E\approx -\bm u\times\bm B$.
To isolate the effect of the fluid velocity $\bm u$ on the magnetic field $\bm B$, we rearrange the terms by moving the first term on the right-hand side to the left-hand side and recall the material derivative $\text{D}/\text{D}t$, which accounts for the change as advected by the fluid. 
This leads to
\begin{equation}
    \DD{\bm B}t=B\nabla_\| \bm u-\bm B\nabla\cdot\bm u.
\end{equation}
Further insight is gained by decomposing $\DD{\bm B}t$ into components parallel and perpendicular to the magnetic field direction $\bm b$.
This decomposition is achieved by expressing $\DD{\bm B}t$ as the sum of $\DD Bt\bm b$ and $B\DD{\bm b}t$, resulting in
\begin{equation}
    \DD{B}t=B\bm{b}\cdot\nabla_\| \bm u-B\nabla\cdot \bm u,
\end{equation}
which describes the change in the magnitude of the magnetic field, and
\begin{align}
    \DD{\bm b}t=(1-\bm{bb})\cdot\nabla_\|\bm u.\label{eq:D11}
\end{align}
Equation (\ref{eq:D11}) is particularly significant as it quantifies the change in the direction of the magnetic field unit vector $\bm b$. 
This equation is instrumental in updating the particles’ parallel momentum, as it captures the effect of the fluid velocity gradient on the orientation of the magnetic field lines, which in turn influences the motion of charged particles along these lines.

\subsection{Equation (\ref{eq:momentum_closure_2})}\label{appendix:eq_F}
To calculate the bulk format of momentum feedback of the gPIC part, we start from decomposing gyro-averaged momentum $\bm P_k$ in to parallel and perpendicular components with respect to the magnetic field, expressed as $\bm P_k=P_{k\|}\bm b+\gamma_k m_k\bm u_\perp$. The time derivative of this momentum is given by
\begin{align}
    \dd{\bm P_k}t&=\dd{P_{k\|}}t\bm b+P_{k\|}\dd{\bm b}t+\dd{\bm \gamma_k}t m_k\bm u_\perp+\gamma_k m_k\dd{\bm u_\perp}t
\end{align}
To account for the variations along the magnetic field line, we use the operator relation $\D/\D t=\text{D}/\text{D}t+\Delta V_{k\|}\nabla_\|$, where $\Delta V_{k\|}\equiv V_{k\|}-u_\|$ represents the difference between the particle’s parallel velocity and the fluid velocity. 
This leads to a more detailed expression %\xb{[Should use capital Vk here and later equations?]}
\begin{equation}
\begin{aligned}
    \dd{\bm P_k}t=&\dd{P_{k\|}}t\bm b+P_{k\|}\DD{\bm b}t+P_{k\|}\Delta V_{k\|}\bm \kappa+\dd{\gamma_k}tm_k\bm u_\perp+\gamma_km_k\DD{\bm u_\perp}t+\gamma_k m_k\Delta V_{k\|}\nabla_\|\bm u_\perp.
\end{aligned}
\end{equation}
To further simplify, we consider the perpendicular component of the fluid velocity derivative
\begin{align}
    \DD{\bm u_\perp}t=\DD{\bm u\cdot(1-\bm{bb})}t=\DD{\bm u}t\bigg|_\perp-\bm u\cdot\DD{\bm b}t\bm b-u_\|\DD{\bm b}t\label{eq:D21}
\end{align}
and the gradient of the perpendicular velocity along the magnetic field line
\begin{align}
    \nabla_\|\bm u_\perp &=\nabla_\|\xkh{\bm u\cdot(1-\bm{bb})}=\DD{\bm b}t-\bm u\cdot\bm{\kappa}\bm b-u_\|\bm \kappa,
\end{align}
where equations (\ref{eq:D11}) and (\ref{eq:electron_momentum}) are applied in deriving above equation. We obtain a comprehensive expression for the time derivative of the particle’s momentum
\begin{equation}
\begin{aligned}
    \dd{\bm P_k}t=&\xkh{q_k E_\|-\frac{\mu_k}{\gamma_k}\nabla_\|B}\bm b+\dd{\gamma_k}tm_k\bm u_\perp+\gamma_k m_k\xkh{\DD{\bm u}t\bigg|_\perp+\Delta V_{k\|}^2\bm\kappa+2\Delta V_{k\|}\DD{\bm b}t},
\end{aligned}
\end{equation}
where we have neglected the perpendicular drift and utilized the approximation
\begin{align}
(\bm V_k-\bm u)\cdot\DD{\bm b}t\approx\Delta V_{k\|}\bm b\cdot\DD{\bm b}t=0.
\end{align}
Summing over all particles and considering the distribution function, we derive the total momentum change within a given volume element
\begin{align*}
    \sum_k\dd{\bm P_k}t\delta(\bm x-\bm X_k)&=\xkh{q_pE_\|-\sfP_{p\perp}\nabla_\|\ln B}\bm b+\xcancel{\frac{\bm u_\perp}{\C^2}\dd{\mathcal E_p}t}+\rho_p\DD{\bm u}t\bigg|_\perp+\sfT_{p\|}\bm \kappa+2\rho_p\Delta u_{p\|}\DD{\bm b}t,
\end{align*}
here, we have employed the definitions from section \ref{sec:fluid_closure}, particularly noting that the parallel component of the stress tensor should be calculated in the gas co-moving coordinate system. 
In our target application, the third term on the right-hand side is negligible due to the small magnitude of the perpendicular velocity relative to the speed of light, $|\bm u_\perp|\ll\C$. 
% \xb{[We also expect to apply this to relativistic ions. Since we need to calculate this term in the energy equation, would it suffice to just copy the results from there?]}
By incorporating the above equation and the expression for the divergence of the perpendicular pressure tensor %\xb{[In the main text, you are using $P_p$]}
% \xb{[Can you elaborate a bit with this derivation?]}
\begin{align}
    \nabla\cdot(\sfP_{p\perp}(1-\bm{bb}))=\nabla_\perp \sfP_{p\perp}-\sfP_{p\perp}\bm\kappa+\sfP_{p\perp}\bm b\nabla_\|\ln B,
\end{align}
into equation (\ref{eq:momentum_closure}), we derive the closed-form equation for momentum conservation, as presented in equation (\ref{eq:momentum_closure_2}).

\subsection{Equation (\ref{eq:energy_closure1})}\label{appendix:eq_W}
To derive the energy closure for gPIC part, we start with the mass-energy equation of particle $k$
\begin{align}
    \xkh{\gamma_km_k\C}^2=m_k^2\C^2+P_{k\|}^2+\xkh{\gamma_km_k\bm u_\perp}^2+2m_k\mu_k B,
\end{align}
the time derivative of the particle’s energy $\varepsilon_k$ provides insight into the energy exchange processes
\begin{align}
    \dd{\varepsilon_k}t&=V_{k\|}\dd{P_{k\|}}t+\gamma_km_k\bm u_\perp\cdot\dd{\bm u_\perp}t+\xcancel{m_ku_\perp^2\dd{\gamma_k}t}+\frac{\mu_k}{\gamma_k}\dd{B}t,
\end{align}
{where we still drop the $\D\gamma_k/\D t$ term on the right-hand side since}
\begin{equation}
    m_ku_\perp^2\dd{\gamma_k}t\ll m_k\C^2\dd{\gamma_k}t=\dd{\varepsilon_k}t.
\end{equation}
%\xb{[This is again for non-relativstic particles. Can you generalize to include this term (then you can only solve the equation implicitly?)]}
Using the momentum equation for electrons (\ref{eq:electron_momentum}), the first term on the right-hand side can be expanded as
\begin{align}
    V_{k\|}\Delta V_{k\|}\bm P_{k}\cdot\bm \kappa+V_{k\|}\underbrace{\bm P_{k}\cdot\DD{\bm b}t}_{\gamma_km_k\bm u_\perp\cdot\DD{\bm b}t}-\frac{\mu_k}{\gamma_k}V_{k\|}\nabla_\|B+qE_\|V_{k\|}.
\end{align}
The second term, involving the perpendicular velocity $\bm u_\perp$, is derived using equation (\ref{eq:D21}),
\begin{equation}
    \gamma_km_k\bm u_\perp\cdot\xkh{\DD{\bm u}t_\perp+(V_{k\|}-2u_\|)\DD{\bm b}t-u_\|\Delta V_{k\|}\bm\kappa}.
\end{equation}
Combining these terms, we arrive at the following expression for the energy change
\begin{equation}
\begin{aligned}
    \dd{\varepsilon_k}t=&\gamma_km_k\bm u_\perp\cdot\xkh{\DD{\bm u}t+2\Delta V_{k\|}\DD{\bm b}t+\Delta V_{k\|}^2\bm\kappa}+q_kE_\|V_{k\|}-\frac{\mu_kB}{\gamma_k}\xkh{\nabla\cdot\bm u_\perp+\bm\kappa\cdot\bm u},
\end{aligned}
\end{equation}
where we have utilized the relationship
\begin{align}
\nabla B\cdot\bm u_\perp+\pp Bt=-B\nabla\cdot\bm u_\perp-B\bm\kappa\cdot\bm u\ .
\end{align}
This leads to the collective energy change for all particles, expressed as 
%\xb{[Again, not straightforward to me.]}
\begin{equation}
\begin{aligned}
    \sum_k\dd{\varepsilon_k}t\delta(\bm x-\bm X_k)=&\bm u_\perp\cdot\xkh{\rho_p\DD{\bm u}t+2\rho_p\Delta u_{p\|}+\sfT_{p\|}\bm\kappa}+E_\|J_{p\|}-\sfP_{p\perp}\xkh{\nabla\cdot\bm u_\perp+\bm\kappa\cdot\bm u}.
\end{aligned}
\end{equation}
By combining this result with the energy closure equation (\ref{eq:energy_closure}), we derive the final form of the energy closure equation (\ref{eq:energy_closure1}).

%% For this sample we use BibTeX plus aasjournalv7.bst to generate the

\bibliography{ref}{}
\bibliographystyle{aasjournalv7}



\end{document}